\documentclass[prd,preprint,superscriptaddress,preprintnumbers,eqsecnum,showpacs,nofootinbib,nobibnotes]{revtex4-1}
\usepackage{amsfonts,amsmath,amssymb,bm}
\allowdisplaybreaks
\usepackage{tikz-feynman}
\usepackage{caption} 
\usepackage{graphicx} 
\usepackage{mathtools}
\usepackage{boondox-cal}
\usepackage{slashed}
\usepackage{hyperref}
\usepackage{xcolor}

\usepackage[utf8]{inputenc}
\usepackage{slashed}

\newcommand{\be}{\begin{equation}}
\newcommand{\bea}{\begin{eqnarray}}
\newcommand{\ba}{\begin{align}}
\newcommand{\ee}{\end{equation}}
\newcommand{\eea}{\end{eqnarray}}
\newcommand{\ea}{\end{align}}

\definecolor{zero2}{rgb}{0.88,0.88,.88}

\def\1eq#1{Eq.~(\ref{#1})}

\def\2eqs#1#2{Eqs.~(\ref{#1}) and~(\ref{#2})}
\def\3eqs#1#2#3{Eqs.~(\ref{#1}),~(\ref{#2}) and~(\ref{#3})}
\def\4eqs#1#2#3#4{Eqs.~(\ref{#1}),~(\ref{#2}),~(\ref{#3}) and~(\ref{#4})}

\def\s#1{{\scriptscriptstyle #1}}

\def\G{\Gamma}

\def\s{\mathcal{s}}

\def\hphi0{{\hat\phi}_0}

\def\d{\!\mathrm{d}^4x\,}

\def\ln{\mathcal{l}}
\def\jn{\mathcal{j}}
\def\lS{\mathcal{l}}
\def\lV{\mathcal{l}'}

\def\Mp2{{\mu}^2}
\def\dEulS{{\cal D}^{M^2}_z}
\def\dEulV{{\cal D}^{\Mp2}_{z'}}

\def\stot{s'}

\def\bpsi{\overline{\psi}}
\def\Ds{\slashed{D}}

\makeatletter
\def\user@resume{resume}
\def\user@intermezzo{intermezzo}
\newcounter{previousequation}
\newcounter{lastsubequation}
\newcounter{savedparentequation}
\def\CT@@do@color{%
	\global\let\CT@do@color\relax
		\@tempdima\wd\z@
		\advance\@tempdima\@tempdimb
		\advance\@tempdima\@tempdimc
		\advance\@tempdimb\tabcolsep
		\advance\@tempdimc\tabcolsep
		\advance\@tempdima1.5\tabcolsep
	\kern-1.5\@tempdimb
	\leaders\vrule
	\hskip\@tempdima\@plus  1fill
	\kern-1.5\@tempdimc
	\hskip-\wd\z@ \@plus -1fill }
\makeatother

\begin{document}

\title{%One-loop
Gauge-Invariant Quantum Fields
}

\date{August 12th, 2024}

\author{A. Quadri}
\email{andrea.quadri@mi.infn.it}
\affiliation{INFN, Sezione di Milano, via Celoria 16, I-20133 Milano, Italy}

\begin{abstract}
\noindent
Gauge-invariant quantum fields are constructed
in an Abelian power-counting renormalizable 
gauge theory with both scalar, vector and fermionic
matter content. 
This extends previous results already obtained for the 
gauge-invariant description of the Higgs mode via
a propagating gauge-invariant field.
The renormalization of the model is studied
in the Algebraic Renormalization approach.

The decomposition of Slavnov-Taylor identities 
into separately invariant sectors is analyzed.

We also comment on some non-renormalizable
extensions of the model whose 1-PI Green's functions
are the flows of certain differential equations of the homogeneous Euler type, exactly resumming the dependence on a certain set of dim. 6 and dim. 8 derivative operators. 

The latter are identified uniquely by the condition that they span
the mass and kinetic terms in the gauge-invariant dynamical fields.
The construction can be extended to non-Abelian gauge groups.
\end{abstract}


\maketitle

\section{Introduction}

While the search of new physics at the LHC
still goes on in the absence of any signals
of new particles, the task of measuring as accurately as possible Standard Model (SM) processes  remains an important goal of both the LHC experiments and  the theoretical effort in the area.

With multi-loops computations nowadays 
routinarily encoded in the phenomenological tools
required 
to study LHC physics, the chiral nature of the SM and the absence of a symmetric (i.e. fulfilling all the relevant functional identities of the theory) regularization scheme pose some technical problems in the evaluation of the symmetry-restoring counter-terms of the Slavnov-Taylor (ST) identities broken by the intermediate regularization (for a recent review see e.g.~\cite{Belusca-Maito:2023wah}).

The problem has been known since a long time~\cite{Ferrari:1999nj,Ferrari:1998jy,Grassi:1999tp,Hollik:1999xh,Hollik:2001cz,Stockinger:2005gx,Fischer:2003cb,Belusca-Maito:2020ala,Cornella:2022hkc}, yet the precision required to confront with experiments in order to assess
evidence of indirect signals of new Beyond-the-Standard-Model (BSM) physics calls for the development of efficient tools in order to ensure the fulfillment of the relevant symmetries of the model and consequently the consistency of the
inserted finite counter-terms.

It might be expected that a control on the separately ST-invariant subsectors of 
one-particle irreducibile (1-PI) amplitudes can simplify the problem, in particular due to the huge number of Feynman diagrams involved in multi-loop computations.

In order to make progress towards this goal,
one can consider to reformulate the theory in terms of dynamical (i.e. propagating inside loops) gauge-invariant composite operators, in one-to-one
correspondence with the elementary fields of the theory.

The procedure must be carried out while preserving
the physical content of the model. The BRST approach is particularly suited to this task,
since it provides a rigorous classification of the contributions to the classical action
that do not alter the physical observables of the theory. Whenever a term
can be written as a BRST-exact functional at zero ghost number 
(i.e. a functional that is the image under the BRST differential $s$ of a functional with ghost number $-1$), the physics is left unchanged. This is what happens for instance for the usual gauge-fixing and ghost terms.

Such a  requirement can be met by introducing  Lagrange multiplier fields, that implement the definition of the gauge-invariant fields, via a suitable
BRST-exact functional. They come together with an appropriate set of additional ghost and antighost fields in order to ensure BRST invariance of the resulting 
classical action.

There are several reasons in favour of such an approach.
Gauge-invariant composite operators have been known since a long time to be the appropriate
tools to describe physics in spontaneously broken gauge models~\cite{Frohlich:1980gj,Frohlich:1981yi,Clark:1974eq}.
For the particular case of the electroweak SU(2) $\times$ U(1) theory, and within the perturbative regime, 
the Fr\"ohlich-Morchio-Strocchi (FMS) construction~\cite{Frohlich:1980gj,Frohlich:1981yi,Maas:2017xzh,Maas:2017wzi} shows that,
for some specific reasons essentially relying on the custodial SU(2) symmetry, states described by gauge-invariant composite operators can be connected to the asymptotic states of elementary W, Z and Higgs modes.
This ensures the reliable standard computation of the S-matrix in terms
of the elementary fields.

In general, however, the equivalence does not hold and 
for a SU(N) gauge group one may even get a different
spectrum in the elementary field description 
as compared with the predictions based on composite gauge-invariant
operators~\cite{Maas:2017xzh,Maas:2017wzi}.

Moreover, the analysis of renormalized
composite gauge-invariant operators
have been recently studied at length 
within the Algebraic Renormalization approach~\cite{Dudal:2023jsu,Dudal:2021dec,Dudal:2021pvw,Dudal:2020uwb,Dudal:2019pyg}, in the formulation where they are coupled to external
classical sources (controlling the renormalization
of such composite operators under loop corrections).

It has been
shown that the gauge-invariant description of the physical Higgs mode via the operator
${\cal O}(x) = \Phi^\dagger \Phi$,
$\Phi$ being the SU(2) Higgs doublet,
consistently leads to a positively
definite spectral density,
unlike in the elementary field 
formalism~\cite{Dudal:2020uwb}.

In the above mentioned approach, the gauge-invariant 
composite operators are treated as
insertions coupled to some suitable
external sources, i.e. one integrates over elementary fields in the path-integral and studies the correlators
of those gauge-invariant operators.

At variance with this approach, one can formulate the theory in such a way that gauge-invariant fields become
dynamical (i.e. propagating inside loops).
This will  hopefully lead to additional relations among 1-PI Green's functions, stemming from the projection of the relevant ST identities in terms of the number of internal gauge-invariant fields. 
Naively, cuts along internal gauge-invariant lines should lead to independent relations, implementing the quartet mechanism~\cite{Kugo:1977zq,Curci:1976yb,Becchi:1974xu}
fulfilling physical unitarity (i.e. the cancellation of unphysical ghost states in physical observables).

This point of view has been already advocated in the
case of the composite operator associated with
the gauge-invariant representation of the Higgs mode~\cite{Binosi:2022ycu}. It has been found that
the ST identities separately hold true
at each order in the number of internal propagators of the gauge-invariant scalar. 

In the present paper we extend the construction
to the case of gauge vector and fermion fields.
We will consider for the sake of definiteness the case of the power-counting renormalizable Abelian Higgs-Kibble model with chiral fermions, but the construction is fully general and can be extended
to non-Abelian gauge group.

The renormalization of the model is studied within the Algebraic Renormalization~\cite{Piguet:1995er} approach. 
We show that  the ST identities can be decomposed into separate sectors according to the number of the internal
propagators of the gauge-invariant variables (either scalars or
vectors).

Gauge-invariant fermion fields cannot be obtained in the model
under investigation, since the charge of the fermionic matter content does not allow to build the gauge-invariant counter-parts of fermions.

This is at variance with the SM, for which the SU(2) left doublets can be mapped into SU(2) gauge-invariant operators, while the SU(2) right singlets are already SU(2) gauge-invariant. The detailed study of the SM formulation in terms of gauge-invariant fields will be presented elsewhere.

The proof of the decomposition of the ST identities into separately invariant sectors can be established in an elegant algebraic way by introducting a suitable set of mass and kinetic terms in the gauge-invariant operators.

These operators in turn are associated with a set of novel differential equations, predicting the exact dependence of 1-PI amplitudes (at fixed order in the loop expansion) on the coefficients of those quadratic operators.

The solutions of such differential equations yield the resummed 1-PI amplitudes in the coefficients of the kinetic and mass operators.
Once mapped to the conventional formalism, the latter turn out to be complicated non-renormalizable interactions, that are difficult to handle in the standard approach, at least if one aims at full resummation in their couplings.

Therefore the gauge-invariant formalism can be thought of as a powerful tool for the study of such classes of BSM models.

\medskip

The paper is organized as follows.
In Sect.~\ref{sec.scalar} we review the construction of gauge-invariant representation of scalar fields. In Sect.~\ref{sec.gauge}
we give the gauge-invariant vector fields both in the Landau and in a generic $R_\xi$-gauge. Fermions are treated
in Sect.~\ref{sec.ferm}. 
In Sect.~\ref{sec:amps} we collect some results on the evaluation of physical observables in comparison with the conventional formalism.
The proof that the ST identity decomposes into separately independent sector is given in Sect.~\ref{sec.sti}.
The diagrammatic interpretation of this decomposition is discussed in Sect.~\ref{sec:diag}.
In Sect.~\ref{sec.pde} we obtain the
partial differential equations
governing the dependence of the 1-PI
amplitudes on the coefficients of the additional
kinetic and mass terms of gauge-invariant vector and scalar fields.
Finally conclusions are 
presented in Sect.~\ref{sec.concl}.
Appendices contain some technical results and the notations used throughout the paper.

%A different question is whether
%one can in general construct suitable
%gauge-invariant variables, 
%corresponding to (combinations of) elementary fields in 
%the linearized approximation,
%over which the path-integral can be carried out directly.
%There are already (partial) examples of 
%such an approach within 
%Effective Field Theory (EFT) approach ({\tt Refs. [28] to [40]
%of https://arxiv.org/pdf/2206.07722.pdf}): in
%the Higgs Effective Field Theory
%the Higgs doublet $\Phi$
%is represented as
%%
%\begin{align}
%    \Phi(x) = 
%    \frac{U(x)}{\sqrt{2}}
%    \begin{pmatrix}
%        0 \cr 
%        v + h(x) \cr
%    \end{pmatrix}
%\end{align}

\section{Scalar field}\label{sec.scalar}

The construction of a dynamical gauge-invariant field for the scalar mode has been presented in~\cite{Quadri:2016wwl,Binosi:2017ubk,Binosi:2019olm,Binosi:2019nwz,Binosi:2020unh,Binosi:2022ycu}. 
The gauge-invariant counterpart of the 
Higgs field is denoted by $h$  and corresponds to the
$X_2$-variable of Ref.~\cite{Binosi:2022ycu} 
(we change the notation  for the scalar gauge-invariant operator in order to set the stage for the generalization to the 
vector and fermion cases):
\begin{align}
    h \sim \frac{1}{v} \Big ( \phi^\dagger \phi - \frac{v^2}{2} \Big ) = \sigma + \dots
    \label{scalar.equiv}
\end{align}
where the dots stand for higher dimensional terms in the fields 
and $\sim$ denotes on-shell equivalence.
$\phi$ is the usual complex Higgs field and $v$ the vacuum expectation value (v.e.v.) of its real part.
We refer the reader to Appendix~\ref{app.cl.act} for our notations and the classical action of the model.

The condition (\ref{scalar.equiv}) is enforced by
introducing a Lagrange multiplier $X$ together with a pair of antighost and ghost fields $\bar c, c$ as follows.
The constraint in Eq.(\ref{scalar.equiv}) is implemented off-shell {\em \`a la} BRST by means of a constraint BRST differential $\s$ such that
\begin{align}
    \s \bar c = h - \frac{1}{v} \Big ( \phi^\dagger \phi - \frac{v^2}{2} \Big ) \, , \quad
    \s X = c \, , \quad \s c = 0 \, , \quad 
    \s h = 0 \, .
\end{align}
Notice that $\s$ is nilpotent due to  
gauge invariance of the r.h.s. of Eq.(\ref{scalar.equiv}) and anticommutes
with the ordinary BRST differential $s$
in Eq.(\ref{gauge.brst}).
Then we add to the action of the 
model in the conventional formalism,
see Eq.(\ref{cl.act}), the following BRST-exact 
term
\begin{align}
S_{\tiny{\mbox{aux,scalar}}} & =
\s \int \d \, \Big [  \bar c (\square + m^2) X \Big ] 
= 
s'  \int \d \, \Big [  \bar c (\square + m^2) X \Big ] 
\nonumber \\
& = 
\int \d \, \Big \{
X (\square + m^2) \Big [ h -  \frac{1}{v} \Big ( \phi^\dagger \phi - \frac{v^2}{2} \Big ) \Big ] 
- \bar c (\square + m^2) c 
\Big \} \, ,
\label{aux.scalar}
\end{align}
i.e.  a contribution that is the image of a functional under the full BRST differential $s' = s + \s$.

This ensures that the physical content
of the theory is not modified~\cite{Barnich:2000zw}.

One also adds to the classical action the quadratic mass term 
\begin{align}
-\frac{M^2-m^2}{2} h^2 \, .
\label{scalar.mass}
\end{align}
This is a physical gauge-invariant mass term for the Higgs mode, as can be seen as follows.
By going on shell with $X$ in Eq.(\ref{aux.scalar}),
one obtains a Klein-Gordon equation
\begin{align}
(\square + m^2) \Big [ h -  \frac{1}{v} \Big ( \phi^\dagger \phi - \frac{v^2}{2} \Big ) \Big ]  = 0 \Rightarrow h = \frac{1}{v}  \Big ( \phi^\dagger \phi - \frac{v^2}{2} \Big ) + \eta \, ,
\end{align}
$\eta$ being a free scalar field with squared mass $m^2$ whose correlators can be proven to vanish
in perturbation theory~\cite{Binosi:2019olm}, so that it can be safely discarded.

Formally this can also be seen
by integrating out in the path integral
the $X$ field and the ghost and antighost fields
$c,\bar c$ in Eq.(\ref{aux.scalar}). Integration over $X$ yields
\begin{align}
    \delta \Big [ (\square + m^2) \Big ( h -  \frac{1}{v} \Big ( \phi^\dagger \phi - \frac{v^2}{2} \Big ) \Big ) \Big ] =
    \frac{1}{| (\square + m^2) \delta_{x,y} |}
    \delta \Big [ h -  \frac{1}{v} \Big ( \phi^\dagger \phi - \frac{v^2}{2} \Big ) \Big ] \, ,
\end{align}
while integration over $c, \bar c$ gives a factor
$(\square + m^2) \delta_{x,y}$ cancelling out the 
denominator in the above equation (up to an inessential multiplicative constant), yielding finally the delta function of the 
constraint in Eq.(\ref{scalar.equiv}).

By substituting back the solution for $h$ into
Eq.(\ref{scalar.mass}), the $m^2$-dependent term cancels against the corresponding contribution in the classical action in Eq.(\ref{cl.act}), leaving finally the Higgs potential
\begin{align}
-\frac{M^2}{2 v^2} \Big (  \phi^\dagger \phi - \frac{v^2}{2} \Big )^2 \, ,
\label{std.higgs.pot}
\end{align}
which shows that the only physical parameter is $M$.

The classical action is invariant under the BRST differential $s$ and under the constraint BRST
differential $\s$.

Several comments are in order here.
First of all we notice that the Lagrange multiplier, enforcing the constraint, only enter via
the BRST-exact contribution $S_{\tiny{\mbox{aux,scalar}}}$. Hence it represents a so-called cohomologically trivial functional~\cite{Barnich:1999cy} and consequently
it cannot alter the physical observables of the model. The latter are identified with the cohomology classes $H(s'|d)$ of the full BRST differential $s' = s + \s$ modulo the exterior derivative $d$
~\cite{Barnich:1999cy,Gomis:1994he,Piguet:1995er}.

At the asymptotic level the additional fields
$X,c$ form a BRST doublet (i.e. a pair of fields one of which is mapped into the other under the BRST differential). BRST doublets drop out of the 
physical spectrum~\cite{Binosi:2022ycu}, to be identified with the Hilbert space ${\cal H}_{\tiny{\mbox{phys}}} = {\mbox{Ker}}~s'_0/ {\mbox{ Im}}~s'_0$,
$s'_0$ being the asymptotic BRST charge of the full BRST differential $s'$.

\section{Gauge field}\label{sec.gauge}

The construction of a dynamical gauge-invariant variable for a massive gauge field 
is more involved, due to the presence of the longitudinal unphysical degree of freedom conspiring with
the pseudo-Goldstone field and the ghosts to ensure the realization of the so-called quartet mechanism guaranteeing the physical unitarity of the theory~\cite{Kugo:1977zq,Curci:1976yb,Becchi:1974xu}.

The relevant gauge-invariant counter-part of the gauge field $A_\mu$ is
\begin{align}
a_\mu & \sim  \frac{i}{e v^2} \Big [ 2 \phi^\dagger  D_\mu \phi - \partial_\mu ( \phi^\dagger \phi ) \Big ] \nonumber \\
& =  \frac{1}{e v^2}
\Big \{  e  [ (v + \sigma)^2 + \chi^2] A_\mu - (v + \sigma) \partial_\mu \chi + \chi \partial_\mu \sigma \Big \} 
= A_\mu - \frac{1}{ev} \partial_\mu \chi + \dots
\label{g.inv.a}
\end{align}
where the dots stand for terms of higher dimension in the fields.

We enforce the on-shell constraint in Eq.(\ref{g.inv.a})
by generalizing 
the procedure presented for the scalar field. 
The additional anti-ghost $\bar c_\mu$ is now a vector field
transforming under the constraint BRST differential $\s$ as
\begin{align}
    \s \bar c_\mu = a_\mu - \frac{i}{ev^2} \Big [ 2 \phi^\dagger  D_\mu \phi - \partial_\mu ( \phi^\dagger \phi ) \Big ] \, , 
\end{align}
while the ghost partner $c_\mu$ pairs into a BRST doublet~\cite{Barnich:2000zw,Gomis:1994he,Quadri:2002nh} with
the Lagrange multiplier $X_\mu$:
\begin{align}
    \s X_\mu = c_\mu \, , \qquad
    \s c_\mu = 0 \, .
\end{align}
Nilpotency of $\s$ follows
by the gauge invariance of the r.h.s. of Eq.(\ref{g.inv.a}). 

The additional terms to be added to the classical vertex functional in order to implement the representation of
the gauge field via the dynamical gauge-invariant variable
$a_\mu$ depend on the specific gauge choice. In fact
we need to add 
\begin{align}
S_{\tiny{\mbox{aux,vect}}} & = \int \d \, \s \Big (
    \bar c_\mu \Sigma_{(\xi)}^{\mu\nu} X_\nu
    \Big ) = 
     \int \d \, s' \Big (
    \bar c_\mu \Sigma_{(\xi)}^{\mu\nu} X_\nu
    \Big )
    \nonumber \\
    & = \int \d \, \Big \{ - \bar c_\mu \Sigma_{(\xi)}^{\mu\nu} c_\nu +
    X_\mu \Sigma_{(\xi)}^{\mu\nu} 
    \Big [ a_\nu - \frac{i}{ev^2} \Big ( 2 \phi^\dagger  D_\nu \phi - \partial_\nu ( \phi^\dagger \phi ) \Big ) \Big ]
    \Big \}\, ,
    \label{X.term}
\end{align}
where the symmetric tensor $\Sigma_{(\xi)}^{\mu\nu}$
generically denotes the 2-point 1-PI amplitude of the gauge field $A_\mu$ in the $R_\xi$-gauge.

We notice that new potentially power-counting violating dim.5 interaction vertices arise from the $X_\mu$-dependent terms
in the second line of Eq.(\ref{X.term}).

The Landau gauge is recovered in the limit $\xi \rightarrow 0$. 

\medskip
(Local) physical observables
of the theory are identified
by the cohomology of the full BRST differential $s'$ in the sector with zero ghost number, i.e.
two operators ${\cal O}, {\cal O}'$ are physically equivalent if they differ by 
a BRST-exact term $s' {\cal R}$,
${\cal O}' = {\cal O} + s' {\cal R}$~\cite{Barnich:2000zw}.

The physical content of the theory is therefore not affected by 
the introduction of $S_{\tiny{\mbox{aux,vect}}}$
in Eq.(\ref{X.term}),
since the latter is a BRST-exact term with ghost number zero.

As in the scalar case, formally this result can be understood by integrating out the field $X_\mu$ and $\bar c_\nu, c_\mu$ in the path integral.
One finds
\begin{align}
\int \, {\cal D} \bar c_\mu
{\cal D} c_\nu
{\cal D} X_\rho
\exp \Big ( i S_{\tiny{\mbox{aux,vect}}} & \Big ) = 
\int \, {\cal D} \bar c_\mu
{\cal D} c_\nu 
\exp \Big ( - i \int \d \bar c_\mu \Sigma_{(\xi)}^{\mu\nu} c_\nu  \Big )
\nonumber \\
&  \times 
\int {\cal D} X_\rho
\exp \Big ( i \int \d \, X_\mu \Sigma_{(\xi)}^{\mu\nu} 
    \Big [ a_\nu - \frac{i}{ev^2} \Big ( 2 \phi^\dagger  D_\nu \phi - \partial_\nu ( \phi^\dagger \phi ) \Big ) \Big ] \Big )
    \nonumber \\
& \sim 
\int \, {\cal D} \bar c_\mu
{\cal D} c_\nu 
\exp \Big ( - i \int \d \bar c_\mu \Sigma_{(\xi)}^{\mu\nu} c_\nu  \Big ) \nonumber \\
& \qquad \times
\delta \Big [ \Sigma_{(\xi)}^{\mu\nu} 
 \Big ( a_\nu - \frac{i}{ev^2} \Big ( 2 \phi^\dagger  D_\nu \phi - \partial_\nu ( \phi^\dagger \phi ) \Big ) \Big ) \Big ]
\nonumber \\
& \sim \det \Sigma_{(\xi)}^{\mu\nu}
\frac{1}{| \det \Sigma_{(\xi)}^{\mu\nu} | }
\delta \Big [  a_\nu - \frac{i}{ev^2} \Big ( 2 \phi^\dagger  D_\nu \phi - \partial_\nu ( \phi^\dagger \phi ) \Big ) \Big ]
\nonumber \\
& \sim \delta \Big [  a_\nu - \frac{i}{ev^2} \Big ( 2 \phi^\dagger  D_\nu \phi - \partial_\nu ( \phi^\dagger \phi ) \Big ) \Big ] \, ,
\label{path.int}
\end{align}
i.e. the determinant arising from the $\delta$-function
is exactly compensated (modulo inessential multiplicative factors) by the integration over the Grassmann variables $\bar c_\mu, c_\mu$.

Several comments are in order here.
From the last line of Eq.(\ref{path.int}) 
one might consider the option to implement the gauge-invariant fields by the Lagrange multiplier technique directly at the level of fields, by eliminating the wave  operator in Eq.(\ref{X.term}) and similarly for
the Klein-Gordon operator in
Eq.(\ref{aux.scalar}).

It turns out that with such a choice the UV behaviour of the fields worstens, making it less transparent to analyze the UV properties of the theory.

On the other hand, if one chooses  higher derivative operators, spurious ghosts are introduced in the model.
The latter  are not easily controlled, calling for specific quantization prescriptions in order to define the theory at the quantum level, like e.g.
the fakeon procedure in
~\cite{Anselmi:2018kgz}.

The choice in Eqs.(\ref{aux.scalar}) and 
(\ref{path.int}) is the simplest one that allows for a diagonalization to mass eigenstates via linear field redefinition and does not involve spurious additional degrees of freedom.

%\det \Sigma_{(\xi)}^{\mu\nu}
\subsection{Landau gauge}

The gauge field propagator in the Landau gauge is transverse and the pseudo-Goldstone field
stay massless. Physical unitarity in this gauge has been studied in detail in Ref.~\cite{Ferrari:2004pd}.

The quadratic part in the relevant sector reads
\begin{align}
    \int \d \Big [ 
    \frac{1}{2} A_\mu ( \square g^{\mu\nu} - \partial^\mu \partial^\nu ) A_\nu + \frac{M_A^2}{2} \Big ( A_\mu - \frac{1}{M_A} \partial_\mu \chi \Big )^2 - b \partial A \Big ] \, ,
    \label{quad.landau}
\end{align}
where $M_A = ev$ is the mass of the gauge boson.
In Landau gauge the relevant operator is
\begin{align}
 \Sigma_{(0)}^{\mu\nu} = 
(\square g^{\mu\nu} - \partial^\mu \partial^\nu) + M_A^2 g^{\mu\nu} \, .
\end{align}
The propagators can be obtained by diagonalizing the 2-point 1-PI amplitudes in the sector spanned by $A_\mu, a_\mu, X_\mu, b, \chi$.
The derivation is presented in Appendix~\ref{app:landau}.

Several comments are in order here.
First one must check that power-counting renormalizability is preserved in the extended field formalism.

This can be seen from the propagators in the symmetric basis 
$(A_\mu, \chi, b, X_\mu, a_\mu)$.
They are given in Eq.(\ref{landau.prop.symm}).
We notice that the $b$-field does not propagate, as a consequence of 
the $b$-equation~(\ref{b.eq}). Moreover,
the propagators $\Delta_{A_\nu A_\nu}$ and $\Delta_{A_\mu a_\nu}$ 
in the first line of Eq.(\ref{landau.prop.symm}) are transverse,
as expected in the Landau gauge. 

Power-counting renormalizability might be spoiled by the
dim. 6 $X_\mu$-dependent interactions in Eq.(\ref{X.term}).
However, we notice that amplitudes with a $X_\mu$-external leg are fixed by the $X_\mu$-equation in Eq.(\ref{Xmu.eq}) in terms of
power-counting renormalizable amplitudes, since the source
$\bar c^{\mu *}$ has positive dimension, see Appendix~\ref{app:fields.dim}.
Thus we only need to consider
amplitudes with an internal $X_\mu$-line in order to assess
the impact of the additional $X_\mu$-dependent dim. 6 interaction vertices in Eq.(\ref{X.term})
on power-counting renormalizability.

We see from Eq.(\ref{landau.prop.symm}) that all propagators involving $X_\mu$, with the exception of $
\Delta_{X_\mu a_\nu}$, are vanishing.
On the other hand, there are no interaction vertices involving
$a_\mu$, so we conclude that 
the additional
dim.6 interactions do not spoil power-counting renormalizability
in the extended field formalism.

Next we observe that the
mass eigenstate $a'_\mu$
in Eq.(\ref{loc.field.redef.2})
is also BRST-invariant,
since by Eq.(\ref{loc.field.redef.0})  it is given by
\begin{align}
a'_\mu = a_\mu - \frac{1}{M_A^2} \partial_\mu b \, ,
\end{align}
i.e. a linear combination of gauge-invariant variables.
Hence one can freely add an independent mass term
\begin{align}
\frac{M_a^2}{2} {a'_\mu}^2
\label{new.mass}
\end{align}
while preserving gauge-invariance.
The $a'_\mu$-propagator is correspondingly modified as 
\begin{align}
    \Delta_{a'_\mu a'_\nu} = \frac{i}{-p^2 + M_A^2 + M_a^2} T_{\mu\nu} + \frac{i}{M_A^2 + M_a^2}L_{\mu\nu} \, . 
\end{align}
The shift in the mass term induces a violation of power-counting renormalizability, since now the $A_\mu$-propagator develops a constant
longitudinal part
\begin{align}
    \Delta_{A_\mu A_\nu} = \frac{i}{-p^2 + M_A^2 + M_a^2} T_{\mu\nu} +
    \frac{ i M_a^2}{M_A^2 ( M_A^2 + M_a^2 )} L_{\mu\nu}
\end{align}
unless $M_a = 0$. 

The violation of power-counting renormalizability by the $a'_\mu$-mass term can be understood by noticing that there are two contributions to the mass term 
\begin{align}
    \int \d \frac{M_a^2}{2} {a'_\mu}^2 & = 
    \int \d \frac{M_a^2}{2} \Big ( a_\mu - \frac{1}{M_A^2} \partial_\mu b 
    \Big )^2 \nonumber \\
    & =  \int \d \Big ( \frac{M_a^2}{2} a_\mu^2 - 
    \frac{M_a^2}{M_A^2}a_\mu \partial^\mu b +
    \frac{M_a^2}{2 M_A^4} \partial^\mu b \partial_\mu b \Big ) \, .
    \label{mass.term.prime}
\end{align}
As a consequence of the gauge invariance of $a_\mu$, 
the last two terms in the above equation can be removed by adding
the BRST-exact term
\begin{align}
    \frac{M_a^2}{M_A^2} \int \d \s \Big [
    \partial_\mu \bar c 
    \Big ( a_\mu - \frac{1}{2 M_A^2}
    \partial_\mu b \Big )
    \Big ] \, . 
    \label{mass.cohom.triv}
\end{align}
and thus they are unphysical.
The first term is on-shell equivalent to the dim.6 operator
\begin{align}
\int \d \, \frac{M_a^2}{2} {a_\mu}^2 \sim
\frac{M_a^2}{2 v^2 M_A^2} \int \d \, 
\phi^\dagger \phi \Big [ 4 (D^\mu \phi)^\dagger D_\mu \phi +
2 \partial^\mu ( \phi^\dagger D_\mu \phi + (D_\mu \phi)^\dagger \phi ) - \square \phi^\dagger \phi
\Big ] \, .
\label{non.ren.term}
\end{align}
The classical action is thus modified by a non-renormalizable interaction.
The relevant term giving a mass contribution to the gauge field
is the first one in the r.h.s.
of Eq.(\ref{non.ren.term}), belonging to the family of operators
\begin{align}
    C_n \equiv \int \d (\phi^\dagger \phi)^n (D^\mu \phi)^\dagger D_\mu \phi \, .
\end{align}
All of them contribute to the
gauge field mass term. As is very well-known, only $C_0$ leads to a
power-counting renormalizable theory.

\subsection{$R_\xi$-gauge}

We now proceed to a generic $R_\xi$-gauge.
The quadratic part in the relevant sector reads
\begin{align}
    \int \d \Big [ 
    \frac{1}{2} A_\mu ( \square g^{\mu\nu} - \partial^\mu \partial^\nu ) A_\nu + \frac{M_A^2}{2} \Big ( A_\mu - \frac{1}{M_A} \partial_\mu \chi \Big )^2 
    + \frac{\xi}{2} b^2 - b ( \partial A + \xi M_A \chi ) \Big ] \, ,
    \label{quad.rxi}
\end{align}
where $M_A = ev$ is again the mass of the gauge boson.

As discussed in Appendix~\ref{app:rxi}, the  operator  $\Sigma_{(\xi)}^{\mu\nu}$ now reads
\begin{align}
 \Sigma_{(\xi)}^{\mu\nu} = 
\Big [ \square g^{\mu\nu} - \Big ( 1 - \frac{1}{\xi} \Big ) \partial^\mu \partial^\nu \Big ] + M_A^2 g^{\mu\nu} \, .
\end{align}
The propagators in the mass eigenstates basis and in the symmetric basis
are reported in Eqs.(\ref{rxi.diag}) and (\ref{rxi.symm}) respectively.

We notice by inspection that the ultraviolet behaviour for large momentum of the propagators
in a $R_\xi$-gauge is the same as in Landau gauge, with the exception
of $\Delta_{X_\mu a_\nu}$, that has a milder behaviour $\sim 1/p^2$ in the $R_\xi$-gauge.
This is due to the lack of subtraction of the longitudinal part described in Landau gauge by $b'$, as can be seen by comparing the last of Eqs.(\ref{landau.diag}) 
with the last of Eqs.(\ref{rxi.diag}).

Since also in a
 $R_\xi$-gauge there are no interaction vertices involving $a_\mu$, the argument ensuring power-counting renormalizability
 in Landau gauge still holds true.

We also notice that $\Delta_{a_\mu a_\nu}$-propagator does not exhibit any pole in the longitudinal part and is the same as in the 
Landau gauge. This is reassuring, since
the gauge-invariant field $a_\mu$ is expected to have
gauge-independent correlator functions and 
its propagator should be that of a massive gauge field without longitudinally propagating degrees of freedom, i.e. a massive gauge field  with the quadratic Proca Lagrangian given by 
$${\cal L}_{\rm Proca} = \frac{1}{2} A_\mu (\square g^{\mu\nu} - \partial^\mu \partial^\nu) A_\nu + \frac{M_A^2}{2} A_\mu^2\, . $$
The Proca Lagrangian is obtained from a generic $R_\xi$-gauge by an operatorial gauge transformation reabsorbing the Goldstone degrees of freedom. The resulting interacting theory is transformed into the so-called unitary gauge.

 We now discuss the effects in the $R_\xi$-gauge of the addition of
 an independent mass term for the gauge field.
 We point out that $a'_\mu$ is a different linear combination now,
 given by the last of Eqs.(\ref{rxi.symm}).
 In particular, it is not gauge invariant.
 Thus we cannot add the mass term in Eq.(\ref{new.mass}), rather one can consider the following mass term:
 \begin{align}
 \frac{M_a^2}{2} a_\mu^2 \, .
 \label{new.mass.xi}
 \end{align}
 Diagonalization proceeds as in Appendix~\ref{app:rxi}
 but the $a_\mu-\chi$-mixing is no more
 removed by the second of Eqs.(\ref{rxi.redef}), so one must invert the $(a_\mu,\chi)$-sector. The resulting propagators are 
\begin{gather}
\Delta_{a_\mu a_\nu} = \frac{i}{-p^2 + M_A^2 + M_a^2}T_{\mu\nu} + \frac{i}{M_A^2 + M_a^2} L_{\mu\nu} \, , \qquad
\Delta_{a_\mu \chi} = -\frac{M_A}{M_A^2 + M_a^2} 
\frac{p^\mu}{-p^2 + \xi M_A^2} \, ,  \nonumber \\
\Delta_{\chi \chi} =
\frac{i M_A^2}{M_A^2 + M_a^2} \frac{ p^2 -
\xi (M_A^2 + M_a^2) }{(-p^2 + \xi M_A^2)^2} \, .
\label{rxi.props}
\end{gather}
Notice that $\Delta_{a_\mu a_\nu}$ does not depend on the gauge parameter, as it should.
The propagator $\Delta_{A_\nu A_\nu}$ equals
$\Delta_{a_\mu a_\nu}$ and still exhibits
a constant longitudinal part, thus destroying power-counting renormalizability.
This is consistent with the fact that
the addition of the new mass term is on-shell equivalent to the introduction of the dim. 6 operator in Eq.(\ref{non.ren.term}).

\section{Fermions}\label{sec.ferm}

According to Eq.(\ref{gauge.brst}) the charge under the U(1) gauge group of the fermionic field $\psi$ is $-\frac{e}{2}$, while the scalar $\phi$ has charge $e$.
Thus it is not possible to construct
a gauge-invariant counter-part of $\psi$ 
 (i.e. a field transformation to a gauge-invariant variable whose linear term coincides with $\psi$) out of $\psi$ and 
the scalar.

This shows that the gauge-invariant variables formalism cannot describe all models with
fermionic matter content.

The SM admits a representation in terms of gauge-invariant variables w.r.t. the SU(2) group.
In fact left-handed leptons are arranged in a SU(2) doublet
\begin{align}
    L \equiv \begin{pmatrix}
        \nu_{eL} \\
        e_L 
    \end{pmatrix} \, , \qquad
    \delta L =  i g \alpha_i \frac{\tau_i}{2} L
\end{align}
$g$ being the SU(2) coupling constant, $\alpha_i$, $i=1,2,3$ the parameters of the infinitesimal SU(2) gauge transformation and 
$\tau_i$ the Pauli matrices.
The right-handed neutrino and electron components $\nu_{eR}$ and $e_R$ are assigned
to the singlet representation of SU(2) and are already gauge-invariant.

On the other hand, the Higgs doublet 
$\Phi = (  \phi^+, \phi^0)^T$ and its charge-conjugated field $\Phi^c \equiv  i \tau_2 \Phi^* = (\phi^{0*}, - \phi^-)^T$
transform both in the fundamental representation of SU(2), i.e.
\begin{align}
\delta \Phi = i g \alpha_i \frac{\tau_i}{2} \Phi \, , \qquad
\delta \Phi^c = i g \alpha_i \frac{\tau_i}{2} \Phi^c \, .
\end{align}
Let us denote by $v$ the vacuum expectation value of $\phi_0$, 
$\langle \Phi \rangle = (0, v)^T$,
$\langle \Phi^c \rangle = (v, 0)^T$.

Thus (the dots denote higher orders terms in the fields)
\begin{align}
    \widetilde e_L \equiv \Phi^\dagger L
    = v e_L + \dots \, , 
    \qquad 
    \widetilde \nu_{eL}  \equiv (\Phi^c)^\dagger  L = v \nu_{eL} + \dots
\end{align}
are both gauge-invariant w.r.t. the SU(2) group and represent
the SU(2) gauge-invariant counter-part of the left-handed electron and the left-handed neutrino respectively.

One should notice that $\widetilde e_L, \widetilde \nu_{eL}$ are not invariant under the hypercharge U(1) gauge group. 
For such fields the Gell Mann-Nishijima representation of the electric charge $Q$ reduces to $Q=Y_W/2$,
$Y_W$ being the hypercharge generator.

We also remark that for quark fields no SU(2) gauge-invariant counter-part can be constructed, due to the charge
mismatch w.r.t. the Higgs doublet.

The detailed analysis of the gauge-invariant formalism for the electroweak theory will be presented elsewhere.

\section{Amplitudes and Physical Observables}\label{sec:amps}

For the sake of convenience we summarize in this Section some results obtained in Refs.~\cite{Binosi:2019olm,Binosi:2019nwz,Binosi:2020unh,Binosi:2022ycu} about amplitudes and physical observables in the 
gauge-invariant field formalism.
\begin{itemize}
\item Computation of physical amplitudes in the gauge-invariant field formalism
is carried out at the level of the gauge-invariant fields. The LSZ reduction formula is used {\em on the mass eigenstates} of the theory, namely one performs the field redefinitions in Eqs.(\ref{scalar.fred}) as well as Eqs. (\ref{landau.vec.fred}) (for the Landau gauge) and (\ref{rxi.vec.fred}) (for the $R_\xi$-gauge).

The relevant asymptotic states are identified
in the usual way by the cohomology of the BRST charge
$Q$ associated with the full BRST differential $\stot = s + \s$. At tree level it is sufficient to consider the linearized approximation of $\stot$. 
The relevant physical Hilbert space is ${\cal H}_{phys} = {\rm Ker }~Q/{\rm Im}~Q$ of asymptotic states that are $Q$-invariant and are not the $Q$-image of another state.

In the scalar sector one has
\begin{align}
[Q, \sigma'] = -c \, , \quad [Q, X']= c \, , \quad [Q,h] = 0 \, ,\quad [Q, \chi] = M_A \omega \, .
\end{align}
In the vector sector one gets:
\begin{enumerate}
\item Landau gauge
\begin{align}
[Q, b'] = -M_A^2 \omega \, , \quad
[Q, A''_\mu] = - c_\mu \, , \quad
[Q, X''_\mu] = c_\mu \, , \quad [Q, a'_\mu ] = 0 \, .
\end{align}
\item $R_\xi$-gauge
\begin{align}
    & [Q, b'] = \frac{1}{\xi} (p^2 - \xi M_A^2) \omega \, ,
    \quad
    [Q, A'_\mu] = - c_\mu - i p_\mu \omega \, , 
    \quad
    [Q, X''_\mu] = c_\mu + i p_\mu \omega \, ,
    \nonumber \\
    & [Q, a'_\mu] = - i p_\mu \omega \, .
\end{align}
\end{enumerate}
In the fermionic sector one has
\begin{align}
    \{ Q, \psi \} = \{ Q, \bar \psi \} = 0 \, .
    \label{Q.fermions}
\end{align}
In the antighost-ghost sector one has
\begin{align}
    & \{ Q, c \} = \{ Q, \omega \} = \{ Q,  c_\mu \} = 0 \, , \quad \{ Q, \bar c \} = - \sigma' - X' \, ,
    \quad  \{ Q, \bar c_\mu \} = -A''_\mu - X''_\mu \, .
    \label{Q.ghosts}
\end{align}
We can therefore spell out the physical states in ${\cal H}_{phys}$ as follows: $\sigma'$ and $X'$ are unphysical since they are not $Q$-invariant; their sum is $Q$-invariant but is the image of the antighost $\bar c$ under $Q$, see Eq.(\ref{Q.ghosts}) and thus it drops out of ${\cal H}_{phys}$. The only physical state is therefore $h$.

Fermions are physical by Eq.(\ref{Q.fermions}).

In the vector sector the enumeration in the Landau gauge goes as follows: $b'$ is unphysical, as well as 
$A''_\mu$ and $X''_\mu$. The combination $A''_\mu + X''_\mu$ is $Q$-invariant but is the $Q$-image of $\bar c_\mu$ by Eq.(\ref{Q.ghosts}), thus it is unphysical.
All the components of $a'_\mu$ are in the kernel of $Q$, however the longitudinal component $\partial a'$ has a constant propagator, see Eq.(\ref{landau.gauge.props}), and therefore is not a physical degree of freedom.
We conclude that the only physical states are the transverse polarizations of the field $a'_\mu$, as it should.

A similar analysis holds true in the $R_\xi$-gauge. $A'_\mu, X''_\mu$ are unphysical by the same argument as before.
On shell at $p^2 = \xi M_A^2$, $b'$ is $Q$-invariant, but again it is not a propagating degree of freedom since its propagator is a constant, see Eq.(\ref{rxi.diag}). 

The 
longitudinal component of the $a'_\mu$-field is not $Q$-invariant, while the transverse ones are. So again we conclude that the only physical states are the transverse polarizations of the field~$a'_\mu$.

The ghost $\omega$ is proportional to the $Q$-variation of $\chi$ 
and thus it is not a physical state. The same reasoning holds true for $c$, that is the $Q$-variation of $X'$.
Furthermore in Landau gauge $c_\mu$ is the $Q$-variation of $X''_\mu$, while in the $R_\xi$-gauge 
it is the $Q$-variation of $X''_\mu+a'_\mu$.
In both cases it is an unphysical degree of freedom.

We finally notice that $\bar \omega$ is also unphysical, since in Landau gauge
\begin{align}
    \{ Q, \bar \omega \} = b' + M_A \chi
\end{align}
while in the $R_\xi$-gauge
\begin{align}
    \{ Q, \bar \omega \} = b' + \frac{1}{\xi} 
    (\partial A' + \partial X'' + \partial a') + M_A \chi
     \, .
\end{align}

Notice that in the mass eigenstate basis
the gauge-invariant fields become interacting and
Eqs.(\ref{ha.eqs.j}) are correspondingly modified, since an
additional functional dependence of the vertex functional on $h, a_\mu$
arises from the field redefinitions in Eqs.(\ref{scalar.fred}) and Eqs. (\ref{landau.vec.fred}) (for the Landau gauge) and (\ref{rxi.vec.fred}) 
(for the $R_\xi$-gauge).

By using the chain rule of differentiation
one gets in the mass eigenstate basis:
\begin{align}
\frac{\delta \G^{(j)}}{\delta h} =  -\frac{\delta \G^{(j)}}{\delta X'} \, , \quad
\frac{\delta \G^{(j)}}{\delta a_\mu} = - \frac{\delta \G^{(j)}}{\delta X''_\mu}+
\frac{\delta \G^{(j)}}{\delta a'_\mu} \, .
\end{align}
\item
The matching with amplitudes in the standard formalism is accomplished according to the following steps:
\begin{enumerate}
\item Radiative corrections are computed in the theory in the gauge-invariant formalism up to the desired order in the loop expansion;
\item One goes on-shell with both the $X$ and $X_\mu$-fields and the $h, a_\mu$ fields.
The resulting 1-PI effective action is a functional in $\phi$ and $A_\mu$ and generates (perturbatively) the 1-PI amplitudes of the ordinary theory. This has been explicitly checked at the  one-loop level in Ref.~\cite{Binosi:2020unh}.
\item Off-shell amplitudes in the gauge-invariant fields differ from their counter-part in the ordinary formalism,
as can be easily seen by computing e.g. the 
four-point connected amplitude of the $h$ field at tree level.
In particular in the gauge-invariant formalism there are momentum-dependent contributions originated by the derivative-dependent $X$- and $X_\mu$-interaction vertices.
However if one considers  a S-matrix element and goes on-shell with the external momenta, the same result is obtained in the conventional and gauge-invariant field formalism.
\end{enumerate}
\end{itemize}

\section{Gauge-invariant sectors and  Slavnov-Taylor identities}\label{sec.sti}

Gauge-invariant quantum fields are associated with a decomposition of the Slavnov-Taylor (ST) identities into separately invariant sectors.

This is a somehow natural consequence of gauge-invariance of the intermediate states described by the gauge-invariant scalar
and vector fields.
In fact, since they are gauge-invariant, they
cannot take part to the quartet mechanism
ensuring the cancellation of unphysical ghost states.

At higher orders both in the SM and, most recently, in 
the program of evaluating radiative corrections in BSM
theories, it is important to check that the ST identities are verified. This is a non-trivial task due to the absence of an invariant regularization scheme for 
chiral theories due to the well known $\gamma_5$-problem.

Dealing with separately invariant sectors may ease
the computation of the finite symmetry-restoring
counter-terms. 
However the correct identification of the separately invariant sectors, involving fermions as well as other fields, is a non-trivial task
at higher orders.
The correct procedure for identifying such sectors is provided by the proposed decomposition, 
  which might consequently be computationally useful.

This is also conceptually interesting since after all the ST identities ensure the cancellation of intermediate unphysical degrees of freedom and gauge-invariant states do not enter into this game, so in principle one should be able to cut internal lines by properly isolating gauge-invariant states and still get combinations of
diagrams that are by themselves ST-invariant.

Intuitevly, this task should accomplished by using 
dynamical gauge-invariant fields inside loops. 
It turns out that this is indeed the case, and the combinatorics involved at the level of Feynman diagrams is  non trivial.

We will discuss this point first at the formal level in
Sect.~\ref{sec.sti} and at the diagrammatic level in Sect.~\ref{sec:diag}.

The algebraic proof of this property is particularly simple and elegant in the Landau gauge.
It extends the technique proposed for the scalar case in
Ref.~\cite{Binosi:2022ycu}.
For that purpose let us add to the classical vertex functional
the quadratic mass operator in Eq.(\ref{new.mass}) 
as well as the contribution to the transverse part of the two-point function 
\begin{align}
    \int \d \, \frac{z'}{2} a'_\mu ( \square g^{\mu\nu} - \partial^\mu \partial^\nu) a'_\nu \, .
    \label{new.kin.term.landau}
\end{align}
One also add the kinetic term for the scalar field proposed in
Ref.~\cite{Binosi:2022ycu}, namely
\begin{align}
    - \int \d \, 
    \frac{z}{2} h \square h \, .
    \label{new.kin.scalar}
\end{align}
We also conveniently choose $M_a^2 = \mu^2 - M_A^2$ in order
to reabsorb the dependence of the gauge field propagator on $M_A^2$.
With this choice the propagators of the $a'_\mu$ and $h$ fields
become
\begin{align}
    \Delta_{hh} = \frac{i}{(1+z)p^2 - M^2} \, , \qquad \Delta_{a'_\mu a'_\nu}(z',\mu^2) = \frac{i}{-(1+z')p^2 + \mu^2} T_{\mu\nu} + \frac{i}{\mu^2}L_{\mu\nu} \, . 
\end{align}
We now rescale the additional parameters  as
\begin{align}
    1 + z \rightarrow \frac{1+z}{t} \, , \quad M^2 \rightarrow \frac{M^2}{t} \, , \quad 1+z' \rightarrow \frac{1+z'}{t'} \, , \quad \mu^2 \rightarrow \frac{\mu^2}{t'} \, .
\end{align}
The rescaling is understood as a book-keeping device in order
to count the number of internal $h$ and $a'_\mu$-propagators.
In fact after the rescaling a 1-PI amplitude $\G^{(n)}_{\varphi_1 \dots \varphi_r}$ at order $n$ in the loop expansion with $r$ external legs 
$\varphi_i = \varphi_i(p_i)$, $i=1, \dots, r$ 
($\varphi_i$ denoting a generic field or external source 
of the theory with incoming momentum $p_i$) can be decomposed
according to
\begin{align}
\G^{(n)}_{\varphi_1 \dots \varphi_r} = \sum_{\ln,\ln'} 
t^\ln {t'}^{\ln'} \G^{(n;\ln,\ln')}_{\varphi_1 \dots \varphi_r} \, .
\end{align}
$\G^{(n;\ln,\ln')}_{\varphi_1 \dots \varphi_r}$ is given by 
 the subset of $n$-loop diagrams with 
$\ln$ internal $h$ lines and $\ln'$ internal
$a'_\mu$ lines, evaluated at $t=t'=1$, plus all the relevant diagrams with lower order counter-terms insertions contributing to that amplitude.

Notice that counter-terms inherit the degree
in $(\ln,\ln')$ induced by their original UV divergent diagrams, e.g. at one loop order a UV divergent diagram with one internal $a'_\mu$-line and one internal $h$ line will give rise to a counter-term
counting as $\ln = 1, \ln '= 1$.

We can now project the ST identity (\ref{sti}) at order $n$ in the loop expansion and at fixed
$(\ln,\ln')$. Thus we  get a tower of identities at order $n$ in the loop expansion
\begin{align}
    \int \d \, \Bigg [ & \partial_\mu \omega 
    \frac{\delta \G^{(n;\ln,\ln')}}{\delta A_\mu} 
    + b \frac{\delta \G^{(n;\ln,\ln')}}{\delta \bar \omega} 
    \nonumber \\
    & + \frac{\delta \G^{(n;\ln,\ln')}}{\delta \sigma^*}
      \frac{\delta \G^{(0)}}{\delta \sigma}
    + \frac{\delta \G^{(0)}}{\delta \sigma^*}
      \frac{\delta \G^{(n;\ln,\ln')}}{\delta \sigma}
     + \frac{\delta \G^{(n;\ln,\ln')}}{\delta \chi^*}
      \frac{\delta \G^{(0)}}{\delta \chi}
    + \frac{\delta \G^{(0)}}{\delta \chi^*}
      \frac{\delta \G^{(n;\ln,\ln')}}{\delta \chi}  
      \nonumber \\
    &  + \frac{\delta \G^{(n;\ln,\ln')}}{\delta \bar \eta}
      \frac{\delta \G^{(0)}}{\delta \psi}
    + \frac{\delta \G^{(0)}}{\delta \bar \eta}
      \frac{\delta \G^{(n;\ln,\ln')}}{\delta \psi}  
     + \frac{\delta \G^{(n;\ln,\ln')}}{\delta  \eta}
      \frac{\delta \G^{(0)}}{\delta \bar \psi}
    + \frac{\delta \G^{(0)}}{\delta  \eta}
      \frac{\delta \G^{(n;\ln,\ln')}}{\delta \bar \psi}  
    \nonumber \\
    & + \sum_{k=1}^{n-1} \sum_{\jn = 0}^\ln 
    \sum_{\jn'=0}^{\ln'} 
    \Big ( 
    \frac{\delta \G^{(k;\jn,\jn')}}{\delta \sigma^*}
    \frac{\delta \G^{(n-k;\ln - \jn,\ln'- \jn')}}{\delta \sigma} + \frac{\delta \G^{(k;\jn,\jn')}}{\delta \chi^*}
    \frac{\delta \G^{(n-k;\ln - \jn,\ln'- \jn')}}{\delta \chi}
    \nonumber \\
    & \qquad \qquad \qquad  +
    \frac{\delta \G^{(k;\jn,\jn')}}{\delta \bar \eta}
    \frac{\delta \G^{(n-k;\ln - \jn,\ln'- \jn')}}{\delta \psi} +
    \frac{\delta \G^{(k;\jn,\jn')}}{\delta \eta}
    \frac{\delta \G^{(n-k;\ln - \jn,\ln'- \jn')}}{\delta \bar \psi} \Big ) 
     \Bigg ] = 0 \, .
     \label{sti.decomp}
\end{align}
After the projection we set
$z=z'=0$ and $\mu^2 = M_A^2$ 
in order to recover the power-counting
renormalizable theory. Yet the decomposition of the ST identities in
Eq.(\ref{sti.decomp}) is more general and applies to the set of non power-counting renormalizable models at arbitrary values
$z, z'$ and $\mu^2$.

The decomposition in Eq.(\ref{sti.decomp})
is particularly interesting. The usual ST identities are in fact relations between 1-PI amplitudes and do not allow to discriminate 
between single subsets of separately invariant diagrams.

On the other hand, Eq.(\ref{sti.decomp})
provides a consistent way to
disentangle those subsets by classifying them
according to the pair $(\ln,\ln')$ given by the number of internal $h$ and $a'_\mu$-lines respectively in the $n$-th order regularized 1-PI amplitude.

\medskip
The same decomposition can be achieved in an arbitrary
$R_\xi$-gauge.

In order to derive the graded ST identity
in Eq.(\ref{sti.decomp}) one needs to diagonalize the propagators of the gauge-invariant variables. 

This can be done in the $R_\xi$-gauge
by going back to Eq.(\ref{rxi.props})
and diagonalizing the quadratic part in
the $(a_\mu,\chi)$-sector by setting
in Fourier space
\begin{align}
    \chi(p) = \chi'(p) - \frac{i M_A}{p^2 - \xi M_A^2} p^\mu a_\mu(p)
    \label{rxi.chi.redef}
\end{align}
The $a_\mu$-propagator in Eq.(\ref{rxi.props}) is unaffected while
the $\chi'$ propagator reads
\begin{align}
\Delta_{\chi'\chi'} =
\frac{- i \xi M_A^2}{(p^2 - \xi M_A^2)^2}
\label{prop.chip.rxi}
\end{align}
At this point one can repeat the argument above by supplementing the theory with
the mass and kinetic terms for the gauge field, carrying out the rescaling and obtain
again Eq.(\ref{sti.decomp}).
Notice that by the redefinition in 
Eq.(\ref{rxi.chi.redef}) new interaction vertices where a $\chi$ has been replaced by $\partial a$ with an attached pseudo-Goldstone propagator will appear.

This is reflected in the fact that the propagator in Eq.(\ref{prop.chip.rxi}) contains a double pole at $p^2 = \xi M_A^2$, so the reshuffling
of the unphysical longitudinal modes among the gauge and the Goldstone fields becomes less transparent than in the Landau gauge.

\section{Diagrammatic interpretation}\label{sec:diag}

In order to elucidate the content of the ST identity
projection in Eq.(\ref{sti.decomp}) it is convenient
to consider the simplest possible case of the 
ST identities for the two-point functions at one loop
order.

For the sake of simplicity and in order to illustrate
the role played by the additional Feynman rules
depending on the extra fields $X,X_\mu$ 
in Eq.(\ref{cl.vertex.functional}) in the easiest situation we work
in the Landau gauge.

We also set $m=0$. This choice switches off the
interaction vertices arising from the quartic potential
in the first line of Eq.(\ref{cl.vertex.functional}).
As has already been checked in Ref.~\cite{Binosi:2020unh}, this does not alter the physical content of the theory while reducing
to the minimum the number of diagrams to be considered.
This will lead us to the cleanest possible example illustrating the (non-trivial) diagrammatic cancellations 
that are summarized by the ST identity projection
in Eq.(\ref{sti.decomp}).

At order one in the loop expansion the ST identity is given by (see Eq.(\ref{sti.full.jth})):
\begin{align}
    {\cal S}_0(\G^{(1)}) = 	
    \int \d  \, \Bigg [ 
    & 
    \partial_\mu \omega 
    \frac{\delta \G^{(1)}}{\delta A_\mu}  + b \frac{\delta \G^{(1)}}{\delta \bar \omega} +
    \frac{\delta \G^{(0)}}{\delta \sigma^*}
    \frac{\delta \G^{(1)}}{\delta \sigma} +
    \frac{\delta \G^{(1)}}{\delta \sigma^*}
    \frac{\delta \G^{(0)}}{\delta \sigma} +
    \frac{\delta \G^{(0)}}{\delta \chi^*}
    \frac{\delta \G^{(1)}}{\delta \chi} +
    \frac{\delta \G^{(1)}}{\delta \chi^*}
    \frac{\delta \G^{(0)}}{\delta \chi}
    \nonumber \\
    & +
    \frac{\delta \G^{(0)}}{\delta \bar \eta}
    \frac{\delta \G^{(1)}}{\delta \psi}
   +
    \frac{\delta \G^{(1)}}{\delta \bar \eta}
    \frac{\delta \G^{(0)}}{\delta \psi} 
    +
    \frac{\delta \G^{(0)}}{\delta \eta}
    \frac{\delta \G^{(1)}}{\delta \bar \psi}    
     +
    \frac{\delta \G^{(1)}}{\delta \eta}
    \frac{\delta \G^{(0)}}{\delta \bar \psi}     
     \Bigg ] = 0 \, .
     \label{sti.oneloop}
\end{align}

The  ST identities for the two-point functions
can be obtained by differentiating Eq.(\ref{sti.oneloop}) 
w.r.t. the ghost $\omega$ and then w.r.t.
$A_\nu$ or the Goldstone field $\chi$ and then setting all the fields and external sources to zero.

In momentum space  we obtain the well-known identities\footnote{In this Section we denote by a subscript $\Phi(p)$ the functional differentiation w.r.t. the field $\Phi$ with incoming momentum $p$, e.g. the two-point gauge 1-PI amplitude is denoted by $\G^{(1)}_{A_\nu(-p) A_\mu(p)} \equiv \frac
{\delta^2 \G^{(1)}}{\delta A_\mu(-p) \delta A_\mu(p)}$.
All fields and external sources are understood to be set to zero after functional differentiation.}:
\begin{align}
& i p^\mu \G^{(1)}_{A_\nu(-p) A_\mu(p)} + \G^{(0)}_{\omega(-p)\chi^*(p)} \G^{(1)}_{A_\nu(-p) \chi(p)} + 
\G^{(1)}_{\omega(-p)\chi^*(p)} \G^{(0)}_{A_\nu(-p) \chi(p)} = 0 \, , \nonumber \\
& i p^\mu \G^{(1)}_{\chi(-p) A_\mu(p)} + \G^{(0)}_{\omega(-p)\chi^*(p)} \G^{(1)}_{\chi(-p) \chi(p)} + 
\G^{(1)}_{\omega(-p)\chi^*(p)} \G^{(0)}_{\chi(-p) \chi(p)} = 0 \, .
\end{align}

One can also take a derivative w.r.t $a_\nu$ and obtain
and additional identity
\begin{align}
& i p^\mu \G^{(1)}_{a_\nu(-p) A_\mu(p)} + \G^{(0)}_{\omega(-p)\chi^*(p)} \G^{(1)}_{a_\nu(-p) \chi(p)} + 
\G^{(1)}_{\omega(-p)\chi^*(p)} \G^{(0)}_{a_\nu(-p) \chi(p)} = 0 \, .
\end{align}
In the Landau gauge the ghosts are free in the Abelian case
and therefore $\G^{(1)}_{\omega(-p)\chi^*(p)}=0$.
Moreover $\G^{(1)}_{\omega(q)\chi^*(p)}=ev \delta^4(q-p)$, thus the ST identities simplify to
\begin{align}
& i p^\mu \G^{(1)}_{A_\nu(-p) A_\mu(p)} + ev \G^{(1)}_{A_\nu(-p) \chi(p)}  = 0 \, , \nonumber \\
& i p^\mu \G^{(1)}_{\chi(-p) A_\mu(p)} + ev \G^{(1)}_{\chi(-p) \chi(p)}  = 0 \, , \nonumber \\
& i p^\mu \G^{(1)}_{a_\nu(-p) A_\mu(p)} + ev \G^{(1)}_{a_\nu(-p) \chi(p)}  = 0 \, .
\label{sti.2pt.1}
\end{align}

Notice that these identities have the same functional form as in the standard formalism, as a consequence of the fact
that the ST identity in the symmetric basis is not modified in the extended field approach. 

Let us start from the first identity. 
The external legs are in the symmetric basis components $(A_\mu, \chi)$. 

Diagrams contributing to the 1-PI two-point functions $\G^{(1)}_{A_\nu(-p) A_\mu(p)}$ are depicted in Figure~\ref{figure:1}. In these diagrams internal propagators are also in the symmetric basis. 

Since $a_\mu$ does not interact and there are no mixed 
$\Delta_{A_\mu X_\nu}$ propagators, the $X_\mu$-dependent interaction vertices
do not appear.

This implies that the gauge propagator in diagram (i) in Fig.~\ref{figure:1} is the standard
gauge propagator $\Delta_{A_\mu A_\nu}$ in Landau gauge and the resulting expression
for $\G^{(1)}_{A_\nu (-p) A_\mu(p)}$ is the same as in the standard formalism.

We report the explicit contributions from the various diagrams in order to ease the comparison for reader's convenience. 
We work in dimensional regularization (DR).
\begin{enumerate}
\item Diagram (i)
\begin{align}
 - 4 e^4 v^2 \int \frac{d^D q}{(2\pi)^D} \, \frac{g^{\mu\nu} - \frac{q^\mu q^\nu}{q^2}}{(q^2 - M_A^2)((p+q)^2 - M^2}
 \end{align} 
\item Diagram (ii)
\begin{align}
e^2 \int \frac{d^D q}{(2\pi)^D} \, \frac{(p^\mu + 2 q^\mu)(p^\nu + 2 q^\nu)}{q^2((p+q)^2-M^2)}
\end{align}
\item Diagram (iii)
\begin{align}
\frac{e^2}{4}  \int \frac{d^D q}{(2\pi)^D} \, \frac{{\rm Tr} [ \gamma^\mu \gamma^5 (\slashed{q} + m_e ) 
    \gamma^\nu \gamma^5 ( \slashed{q} + \slashed{p} + m_e )]}{(q^2 - m_e^2)( (q+p)^2 - m_e^2 )}
\end{align}
\item Diagram (iv)
\begin{align}
- e^2  \int \frac{d^D q}{(2\pi)^D} \, \frac{g^{\mu\nu}}{q^2 - M^2} 
\end{align}
\item Diagram (v)

It vanishes in DR.
\end{enumerate}

\begin{figure}[h]
\centering
\begin{tabular}{ccc}  % Adjust the number of columns as needed using 'c', 'l', or 'r'

% First diagram
\begin{minipage}{.3\textwidth}
  \centering
  \feynmandiagram [layered layout, horizontal=b to c] {
    a  -- [scalar] b
    -- [photon, half left, looseness=1.5] c
    -- [half left, looseness=1.5] b,
    c  -- [photon] d ,
    };
\end{minipage}
&  % Column separator

% Second diagram
\begin{minipage}{.3\textwidth}
  \centering
  \feynmandiagram [layered layout, horizontal=b to c, horizontal=b to d] {
    a  -- [scalar] b
    -- [scalar,  half left, looseness=1.5] c
    -- [color=red, half left, looseness=1.5] b,
    c  --  [photon] d,
    };
\end{minipage}

&  % Column separator

% Third diagram
\begin{minipage}{.3\textwidth}
  \centering
  \feynmandiagram [layered layout, horizontal=b to c, horizontal=b to d] {
    a  -- [scalar] b
    -- [fermion, half left, looseness=1.5] c
    -- [fermion, half left, looseness=1.5] b,
    c  -- [photon] d ,
    };
\end{minipage}
\cr

\begin{minipage}{.3\textwidth}
  \centering
  (I)
\end{minipage}

&

\begin{minipage}{.3\textwidth}
  \centering
  (II)
\end{minipage}

&

\begin{minipage}{.3\textwidth}
  \centering
  (III)
\end{minipage}

\end{tabular}
\caption{
{\small 
$\G^{(1)}_{\chi A_\mu}$. Internal lines are in the symmetric basis. The red line denote the propagator~$\Delta_{\sigma X}$.
}
}
\label{figure:2}
\end{figure}
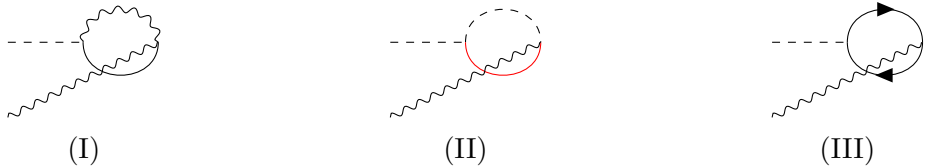

Diagrams contributing to the amplitude $\G^{(1)}_{A_\nu(-p) \chi(p)}$ are represented in Figure~\ref{figure:2}.
Their expressions are given by:
\begin{enumerate}
\item Diagram (I)
\begin{align}
- 2 i e^3  v  \int \frac{d^D q}{(2\pi)^D} \, \frac{( 2 p_\nu + q_\nu) \Big ( g^{\mu\nu} - \frac{q^\mu q^\nu}{q^2} \Big )}{(q^2 - M_A^2)((p+q)^2 - M^2)} 
\end{align}
This amplitude coincides with the corresponding one in the conventional formalism.
\item Diagram (II)

Diagram (II) is  generated by the insertion of the $X$-dependent vertex
$- \frac{1}{2v} \square X \chi^2$, arising from the third line of 
Eq.(\ref{cl.vertex.functional}).

This interaction vertex is not present in the conventional formalism. 
In order to understand how the correspondence with the conventional amplitude is recovered,
 we notice that the mixed $\Delta_{X\sigma}$ propagator does not vanish and is given in the case $m=0$
 by (see Eq.(\ref{scalar.symm.props}))
$$\Delta_{X\sigma}(q) = - \frac{i M^2}{(q^2 - M^2)q^2}\, .$$

Moreover
\begin{align}
\G^{(0)}_{X(-q) \chi(-p) \chi(p+q)} \Delta_{X\sigma}(q)  = 
\frac{q^2 }{v} \frac{-i M^2}{(q^2 - M^2)q^2} = \frac{- M^2}{v} \frac{i}{q^2-M^2} \, , 
\end{align}
i.e. the $q^2$-factor from the derivative-dependent interaction term  
cancels out the $q^2$-dependence in the denominator of
$\Delta_{X\sigma}(q)$, leaving an effective vertex
\begin{align}
V_{\sigma \chi^2} = -\frac{M^2}{v} \, .
\end{align}
This is the usual interaction vertex generated by the standard quartic potential 
Eq.(\ref{std.higgs.pot}),
%$-\frac{M^2}{2v^2} \Big ( \phi^\dagger \phi - \frac{v^2}{2} \Big )$, 
proportional to the  mass of the physical scalar $M$.

The amplitude associated with diagram (II) is finally given by
\begin{align}
\frac{i e M^2}{v}  \int \frac{d^D q}{(2\pi)^D} \, \frac{(p^\mu + 2 q^\mu) }{(q^2 - M^2)(p+q)^2} 
\end{align}
i.e. the same expression as in the standard formalism.

A comment is in order here. 

If $m$ were different than zero, another diagram should be included. It is generated by the additional trilinear $\sigma\chi^2$ vertex proportional to $m$ from the quartic Higgs potential in the first line of Eq.(\ref{cl.vertex.functional}). 
The diagram with the exchange of the $\Delta_{X\sigma}$-propagator would be now $\Delta_{X\sigma}(q) = -\frac{i (M^2 - m^2)}{(q^2 - M^2)(q^2-m^2)}$ (see Eq.(\ref{scalar.symm.props}))
and the $X$-dependent interaction vertex from the third
line of Eq.(\ref{cl.vertex.functional})  would  provide a factor $q^2 -m^2$ in the numerator of the loop amplitude,
cancelling against its inverse in the $\Delta_{X\sigma}$-propagator. 

The net result is still a contribution proportional to $\frac{M^2 - m^2}{q^2 - M^2}$. The $m^2$-piece of the amplitude cancels against the diagram from the usual quartic Higgs potential still from the first line of
Eq.(\ref{cl.vertex.functional}), thus leaving the only
$M^2$-piece.

This is an example of the cancellation of the $m^2$-dependence
that has been discussed diagrammatically in Ref.~\cite{Binosi:2019nwz}. It is a consequence of the fact that $m^2$ is not a physical parameter of the theory, as can be easily seen on general grounds by noticing that the whole dependence of the classical action on $m^2$ is confined to a BRST-exact term, see Eq.(\ref{aux.scalar}).

\item Diagram (III)
\begin{align}
i \frac{ G e}{2} \int \frac{d^D q}{(2\pi)^D} \frac{{\rm Tr} [  \gamma^5 (\slashed{q} + m_e ) 
    \gamma^\mu \gamma^5 ( \slashed{q} + \slashed{p} + m_e )]}{(q^2 - m_e^2)( (q+p)^2 - m_e^2 )}
\end{align}
The diagram coincides with the one in the conventional formalism.
\end{enumerate}

\begin{figure}[h]
\centering
\begin{tabular}{ccc}  % Adjust the number of columns as needed using 'c', 'l', or 'r'

% First diagram
\begin{minipage}{.3\textwidth}
  \centering
  \feynmandiagram [layered layout, horizontal=b to c, horizontal=b to d] {
    a  -- [photon] b
    -- [photon, half left, looseness=1.5] c
    -- [half left, looseness=1.5] b,
    c  -- [photon] d,
    };
    (i)
\end{minipage}

&  % Column separator

% Second diagram
\begin{minipage}{.3\textwidth}
  \centering
  \feynmandiagram [layered layout, horizontal=b to c, horizontal=b to d] {
    a  -- [photon] b
    -- [scalar, half left, looseness=1.5] c
    -- [half left, looseness=1.5] b,
    c  --  [photon] d,
    };
    (ii)
\end{minipage}
&  % Column separator

%Third diagram
\begin{minipage}{.3\textwidth}
  \centering
  \feynmandiagram [layered layout, horizontal=b to c, horizontal=b to d] {
    a  -- [photon] b
    -- [fermion, half left, looseness=1.5] c
    -- [fermion, half left, looseness=1.5] b,
    c  --  [photon] d,
    };
    (iii)
\end{minipage}

% Column separator
\cr

%First diagram / 2nd line
\begin{minipage}{.3\textwidth}
  \centering
   \feynmandiagram [layered layout, horizontal=a to c, horizontal=b to d] {
    a -- [photon] b -- [out=135, in=45, loop, min distance=2cm] b,
    b -- [photon] c,
    };
\end{minipage}

&  % Column separator

% Second diagram / 2nd line
\begin{minipage}{.3\textwidth}
  \centering
   \feynmandiagram [layered layout, horizontal=a to c, horizontal=b to d] {
    a -- [photon] b -- [scalar, out=135, in=45, loop, min distance=2cm] b,
    b -- [photon] c,
    };
\end{minipage}
&
% Third diagram / 2nd line
\begin{minipage}{.3\textwidth}
 \end{minipage}

\cr
\begin{minipage}{.3\textwidth}
\centering
(iv)
\end{minipage}
&
\begin{minipage}{.3\textwidth}
\centering
(v)
\end{minipage}

&
\begin{minipage}{.3\textwidth}
\centering
\end{minipage}

\end{tabular}
\caption{
{\small $\G^{(1)}_{A_\nu A_\mu}$. Internal lines are in the symmetric basis. The wavy line denotes the $\Delta_{A_\mu A_\nu}$-propagator, the solid line the $\Delta_{\sigma\sigma}$-propagator while the dashed line is the Goldstone
$\Delta_{\chi\chi}$-propagator. The last diagram in the first row is the fermion loop.}
}
\label{figure:1}
\end{figure}
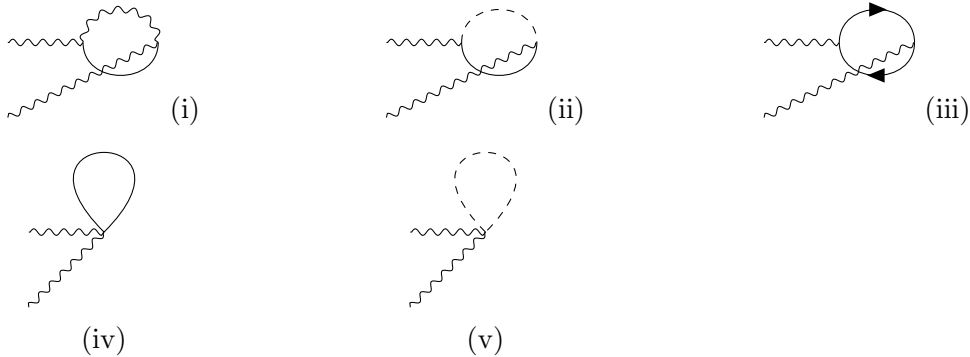

By explicit computation one can easily check that the first of the ST identities in Eq.(\ref{sti.2pt.1}) is satisfied. This comes as no suprise, since one can easily map the contributions from the diagrams in the auxiliary field formalism to the standard ones,
as has been shown above.

\subsection{Invariant subsectors}

Then one can ask oneself whether this identity can in fact be split into separately invariant subsectors. This is motivated by the fact that, for instance, the fermionic sector  is invariant  by itself~\footnote{Since no invariant 
regularization scheme is known in the presence of the $\gamma_5$ matrix, finite counter-terms 
must be added in order to recover the ST invariance broken by intermediate regularization.}, so one might wonder whether one can rearrange  contributions to the individual diagrams in such a way that  they exhibit independently the ST invariance.

Intuitively, this procedure should in principle be possible: the ST identities control the cancellation among unphysical degrees of freedom in intermediate states via the so-called quartet mechanism~\cite{Becchi:1974xu,Curci:1976yb,Kugo:1977zq}, while gauge invariant modes in the intermediate states do not enter such cancellations. Therefore, diagrams with the exchange of the same set of internal gauge-invariant states should cluster together into independently invariant ST sectors.

For that purpose one needs to look at the gauge-invariant
variables propagating inside the loop.
More precisely,  each internal propagator entering 
in the diagrams in Figures~\ref{figure:1},\ref{figure:2} 
must be re-expressed as a sum over propagators
in the mass eigenstate basis according to 
the decomposition in Eq.(\ref{landau.diag}).
For instance, the propagator
$\Delta_{A_\mu A_\nu}$ is decomposed according to
\begin{align}
\Delta_{A_\mu(-p) A_\nu(p)} = \Delta_{A''_\mu(-p) A''_\nu(p)} +  \Delta_{X''_\mu(-p) X''_\nu(p)} + \frac{1}{M_A^4} p^\mu p^\nu \Delta_{b'(-p) b'(p)} + 
\Delta_{a'_\mu(-p) a'_\nu(p)}
 \, ,
\end{align}
in agreement with the second equation in the first line
of Eq.(\ref{landau.diag}) and similarly for the other fields.
It has a zero-th order component in the number of internal $a'_\mu$-lines, given by the first three terms in the r.h.s. of the above equation, and a first order component in the number of internal $a'_\mu$-lines, given by the last term in the above equation. Obviously it has no component at first order in the number of internal $h$-lines. Hence
\begin{align}
& \Delta^{(0,0)}_{A_\mu(-p) A_\nu(p)} = \Delta_{A''_\mu(-p) A''_\nu(p)} +  \Delta_{X''_\mu(-p) X''_\nu(p)} + \frac{1}{M_A^4} p^\mu p^\nu \Delta_{b'(-p) b'(p)} = -\frac{i}{M_A^2}
\frac{p^\mu p^\nu}{p^2} \, , \nonumber \\
& \Delta^{(0,1)}_{A_\mu(-p) A_\nu(p)} = \Delta_{a'_\mu(-p) a'_\nu(p)} \, , \qquad \Delta^{(1,i)}_{A_\mu(-p) A_\nu(p)} = 0 \, , ~~~ i=0, 1 .
\end{align}
Similarly one finds that $\Delta_{\chi\chi}$ is of order $(0,0)$, while 
\begin{align}
   & \Delta^{(0,0)}_{XX} = \Delta_{X'X'} = -\frac{i}{p^2} \, , \quad
     \Delta^{(1,0)}_{XX} = \Delta_{hh} \, , \quad 
     \Delta^{(0,0)}_{\sigma\sigma} = \Delta_{\sigma'\sigma'}+\Delta_{X'X'} = 0 \, , \quad
     \Delta^{(1,0)}_{\sigma\sigma} = \Delta_{hh} \, ,
     \nonumber \\
   & \Delta^{(0,0)}_{X\sigma} = -\Delta_{X'X'} \, , \qquad
    \Delta^{(1,0)}_{X\sigma} = -\Delta_{hh}  \, , \qquad
   \Delta^{(i,1)}_{XX} = \Delta^{(i,1)}_{X\sigma} = \Delta^{(i,1)}_{\sigma\sigma} = 0 \, , ~~i = 0,1 \, .
\end{align}
The resulting diagrams can correspondingly be rearranged according to a double index $(\ln,\ln')$ in the number of internal
mass eigenstates lines for the field $h$ and for the
field $a'_\mu$ respectively. Notice that in the mass eigenstate basis both $h$ and $a'_\mu$ do interact with other fields, so the decomposition is non-trivial.

Moreover, in the sectors $(\ln,\ln')$ it is no more true
that the $X_\mu$-propagators vanish.
In particular one has from Eq.(\ref{scalar.symm.props})
\begin{align}
& \Delta_{X_\mu X_\nu}^{(0,0)} = \Delta_{X_\mu A_\nu}^{(0,0)} =\Delta_{X''_\mu X''_\nu} \, ,
\quad
\Delta_{X_\mu X_\nu}^{(0,1)} =  \Delta_{X_\mu A_\nu}^{(0,1)}= \Delta_{a'_\mu a'_\nu} \, ,
\quad
\Delta_{X_\mu X_\nu}^{(1,i)} = \Delta_{X_\mu A_\nu}^{(1,i)} = 0, ~~ i=0,1 \, .
\label{prop.XX.XA}
\end{align}
%%%%%%%p
But since the $(\ln,\ln')$-components of the $X_\mu$-propagators do not vanish, in the sectors $(\ln,\ln')$
diagrams involving the $X_\mu$-dependent interaction vertices in the fourth line of Eq.(\ref{cl.vertex.functional}) become crucial.

We analyze separately the ST identities projection on the different sectors.
This amounts: i) to project the diagrams in Fig.~\ref{figure:1} and Fig.~\ref{figure:2} on the relevant sector; ii) to add the contributions in  that particular sector from the additional diagrams involving the $X_\mu$-propagators.  They are depicted in Figs.~\ref{figure:3} and ~\ref{figure:4}.

As an example, let us work out in detail the projections of diagram (i) in Fig.~\ref{figure:1}.
At order $(0,0)$ one has
\begin{align}
-  4 e^4 v^2 \int \frac{d^Dq}{(2\pi)^D}  \Delta^{(0,0)}_{\sigma\sigma}(p+q) \Delta^{(0,0)}_{A_\mu A_\nu}(q)
\end{align}
which vanishes identically since $\Delta^{(0,0)}_{\sigma\sigma}=0$.

At order $(0,1)$ the projection 
\begin{align}
-  4 e^4 v^2 \int \frac{d^Dq}{(2\pi)^D}  \Delta^{(0,0)}_{\sigma\sigma}(p+q) \Delta^{(1,0)}_{A_\mu A_\nu}(q)
\end{align}
similarly vanishes for the same reason.

On the other hand at orders $(1,0)$ and $(1,1)$  one gets non-trivial results, since
\begin{align}
& -  4 e^4 v^2 \int \frac{d^Dq}{(2\pi)^D}  \Delta^{(1,0)}_{\sigma\sigma}(p+q) \Delta^{(0,0)}_{A_\mu A_\nu}(q)
= -\frac{4 e^4 v^2}{M_A^2}  \int \frac{d^Dq}{(2\pi)^D}  \frac{q^\mu q^\nu}{q^2 ( (q+p)^2 - M^2 )} \, , \nonumber \\
& -  4 e^4 v^2 \int \frac{d^Dq}{(2\pi)^D}  \Delta^{(1,0)}_{\sigma\sigma}(p+q) \Delta^{(0,1)}_{A_\mu A_\nu}(q)
\nonumber \\
& \qquad \qquad \qquad 
= 4 e^4 v^2  \int \frac{d^Dq}{(2\pi)^D}  \frac{1}{ (q+p)^2 - M^2} 
 \Big [ 
-\frac{1}{q^2-M_A^2} \Big ( g^{\mu\nu} - \frac{q^\mu q^\nu}{q^2} \Big ) +
\frac{1}{M_A^2} \frac{q^\mu q^\nu}{q^2} \Big ] \, .
\end{align}
Let us now discuss in detail the separate cancellations happening in each sector.
\begin{itemize}
\item $(0,0)$-sector

In this sector there are no contributions from
diagrams (i), (ii) and (iv) in Fig.~\ref{figure:1} since $\Delta_{\sigma\sigma}^{(0,0)}$ vanishes.  
Diagram (I) in Fig.~\ref{figure:2}  vanishes by the same reason.
Diagram (v) vanishes in DR, diagram (II) yields
$$
i \frac{e}{v} \int \frac{d^D q}{(2\pi)^D} \frac{q^2 (p^\mu + 2 q^\mu)}{q^2 ( p+q)^2} 
$$
and again the $q^2$ from the interaction vertex $\G^{(0)}_{X\chi\chi}$ cancels against the pole from the 
$\Delta_{\sigma X}^{(0,0)}$ propagator. The resulting integral is zero in DR.

The final result is that the ST projection on the $(0,0)$-sector gives the separate ST invariance of the fermionic sector,
diagram  (iii) in Figure~\ref{figure:1} and diagram (III) in Figure~\ref{figure:2}.
\item $(0,1)$-sector

In this sector (zero internal $h$ lines and one internal $a'_\mu$ line) the potential contribution is only from
diagram (i) in Fig.~\ref{figure:1} and from
diagram (I) in Fig.~\ref{figure:2} when the upper
vector propagator is taken at order $1$ in the
gauge-invariant field $a'_\mu$. However 
$\Delta_{\sigma\sigma}$ at order zero in the $h$-lines vanishes and therefore the ST identities projection 
at order $(0,1)$ vanishes identically.
\item $(1,0)$-sector

The $(1,0)$-projection of the ST identities is non-trivial.
The relevant contributions are the following:
\begin{enumerate}
\item diagram (i) in Fig.~\ref{figure:1} contributes with an upper vector propagator $\Delta_{A_\mu A_\nu}^{(0,0)}$ and
a lower scalar propagator $\Delta_{\sigma\sigma}^{(1,0)}$:
\begin{align}
{\cal A}_{\mu\nu;(i)}^{(1,0)}(p) = - 4 e^4 v^2 \int \frac{d^D q}{(2\pi)^D} \Delta_{\sigma\sigma}^{(1,0)}(p+q) \Delta_{A_\mu A_\nu}^{(0,0)}(q) \,;
\end{align}
\item similarly diagrams (ii) and (iv) in Fig.~\ref{figure:1} contribute with a scalar propagator $\Delta_{\sigma\sigma}^{(1,0)}$:
\begin{align}
{\cal A}_{\mu\nu;(ii)}^{(1,0)}(p) =  - i e^2  \int \frac{d^D q}{(2\pi)^D} \frac{(p_\mu + 2 q_\mu)(p_\nu + 2 q_\nu)}{q^2}
\Delta_{\sigma\sigma}^{(1,0)}(p+q) \,;
\end{align}
\begin{align}
{\cal A}_{\mu\nu;(iv)}^{(1,0)}(p) =  i e^2 g_{\mu\nu}  \int \frac{d^D q}{(2\pi)^D} 
\Delta_{\sigma\sigma}^{(1,0)}(q) \,;
\end{align}
\item there are no contributions from diagrams (iii) and (v);
\item diagram (I) in Fig.~\ref{figure:2} contributes with an upper vector propagator $\Delta_{A_\mu A_\nu}^{(0,0)}$ and
a lower scalar propagator $\Delta_{\sigma\sigma}^{(1,0)}$:
\begin{align}
{\cal A}_{\nu;(I)}^{(1,0)}(p) =  2 i e^3 v  \int \frac{d^D q}{(2\pi)^D} (2 p_\mu + q_\mu) \Delta_{\sigma\sigma}^{(1,0)}(p+q) \Delta_{A_\mu A_\nu}^{(0,0)}(q) \,;
\end{align}
\item diagram (II) in Fig.~\ref{figure:2} contributes 
with a propagator $\Delta^{(1,0)}_{X\sigma}$:
\begin{align}
{\cal A}_{\nu;(II)}^{(1,0)}(p) =  -\frac{ e}{v}  \int \frac{d^D q}{(2\pi)^D} q^2 (p_\nu + 2 q_\nu) \Delta_{X\sigma}^{(1,0)}(q) 
\frac{1}{(p+q)^2} \,.
\end{align}
\end{enumerate}
Their explicit expressions are reported in Appendix~\ref{app:2pt.sti}.

By explicit computation one sees that these diagrams do not satisfy by themselves the ST identities.

In order to obtain the correct projection on the 
sector $(1,0)$, some extra Feynman diagrams must be added.
They involve the $X_\mu$-dependent interaction vertices
and the $X_\mu$-propagators that are precisely those predicted by the 
Feynman rules in Eq.(\ref{cl.vertex.functional}).

The relevant interaction vertex is the following trilinear coupling from the fourth line of Eq.(\ref{cl.vertex.functional}):
\begin{align}
\G^{(0)}_{X_\mu(q) A_\nu \sigma} = -\frac{2}{v} \Sigma^{\mu\nu}_{(0)}(q) \, . 
\label{X.vert.1}
\end{align}
It gives rise to the diagram $(\alpha)$ in Fig.~\ref{figure:3} (the red wavy line is here the propagator $\Delta_{X_\mu X_\nu}^{(0,0)}$)
and to the diagram $(\beta)$ in the same Figure:
\begin{align}
& {\cal A}_{\mu\nu;(\alpha)}^{(1,0)}(p) = - \frac{4}{v^2}  \int \frac{d^D q}{(2\pi)^D} \Sigma^{\mu\alpha}_{(0)}(q)\Sigma^{\nu\beta}_{(0)}(q)
\Delta_{\sigma\sigma}^{(1,0)}(p+q) \Delta_{X_\alpha X_\beta}^{(0,0)}(q) \, ; \nonumber \\
& {\cal A}_{\mu\nu;(\beta)}^{(1,0)}(p) = 8 e^2  \int \frac{d^D q}{(2\pi)^D} \Sigma^{\mu\alpha}_{(0)}(q)
\Delta_{\sigma\sigma}^{(1,0)}(p+q) \Delta_{X_\alpha A_\nu}^{(0,0)}(q) \, .
\end{align}

 In the latter case the vector propagator is $\Delta_{X_\mu A_\nu}^{(0,0)}$
and the $A_\nu$-line enters into the usual trilinear vertex $A^2\sigma$ from the covariant kinetic term.

The contributions to $\G^{(1)}_{A_\nu \chi}$ are depicted in Fig.~\ref{figure:4}. They are generated by the interaction term
\begin{align}
\G^{(0)}_{X_\mu(-(p+q)) \sigma(q) \chi(p)} = \frac{i}{ev^2}  \Sigma^{\mu\nu}_{(0)}(-(p+q)) (q_\mu - p_\mu) 
\label{X.vert.2}
\end{align}
where again in the $(1,0)$-sector the propagators $\Delta_{X_\mu X_\nu}^{(0,0)}$ and $\Delta_{X_\mu A_\nu}^{(0,0)}$ enter in the loop:
\begin{align}
&
{\cal A}_{\nu;(\gamma)}^{(1,0)}(p) =  \frac{2 i}{e v^3}  \int \frac{d^D q}{(2\pi)^D} ( 2 p^\rho + q^\rho) \Sigma^{\nu\alpha}_{(0)}(q)
\Sigma^{\beta\rho}_{(0)}(q) \Delta^{(0,0)}_{X_\alpha X_\beta}(q) \Delta^{(1,0)}_{\sigma\sigma}(k+q) \, ; \nonumber \\
& {\cal A}_{\nu;(\delta)}^{(1,0)}(p) =  - \frac{2 i e}{v}  \int \frac{d^D q}{(2\pi)^D}  \Big [ (2 p^\alpha + q^\alpha)\Sigma^{\nu\beta}_{(0)}(q) \Delta_{X_\alpha A_\beta}^{(0,0)}(q)\Delta^{(1,0)}_{\sigma\sigma}(k+q)  \nonumber \\
& \qquad \qquad \qquad
+ (2 p^\beta + q^\beta)\Sigma^{\alpha\beta}_{(0)}(q) \Delta_{X_\alpha A_\nu}^{(0,0)}(q)
\Delta^{(1,0)}_{\sigma\sigma}(k+q)  \Big ] \, .
\end{align}

The corresponding contributions are listed in Eq.(\ref{xmu.contr}).
%%%%

In order to see how the cancellation mechanism works, we notice that both vertices (\ref{X.vert.1}) and (\ref{X.vert.2}) 
involve the operator $\Sigma^{\mu\nu}_{(0)}$ acting on the propagators
$\Delta^{(0,0)}_{X_\mu X_\nu} = \Delta_{X''_\mu X''_\nu}$ and $\Delta^{(0,0)}_{X_\mu A_\nu} = \Delta_{X''_\mu X''_\nu}$.

We also notice the following identities:
\begin{align}
\Sigma^{\mu\alpha}_{(0)}(q)  \Delta_{X''_\alpha X''_\nu}(q^2) = - i g^\mu_\nu \, , \qquad
\Sigma^{\mu\alpha}_{(0)}(q)  \Delta_{X''_\alpha X''_\beta}(q^2) \Sigma^{\beta\nu}_{(0)}(q) = - i \Sigma^{\mu\nu}_{(0)}(q) \, . 
\end{align}
Therefore the pole in $q^2=M_A^2$ present in $\Delta_{X''_\mu X''_\nu}$ is removed by the
projection against $\Sigma^{\mu\nu}_{(0)}$.
One is left with Feynman amplitudes whose integrands only exhibit a pole at $q^2=M^2$ from
the scalar propagator $\Delta^{(1,0)}_{\sigma\sigma}(q^2)$, i.e. the analytic structure obtained 
from diagram (i) once the vector propagator is replaced by $\Delta^{(0,0)}_{A_\mu A_\nu}$.

The relative coefficients of the various diagrams are dictated by the extra
$X_\mu$-dependent Feynman rules in Eq.(\ref{cl.vertex.functional}).

By summing all the diagrams one obtains the following ST identity projected in the $(1,0)$-sector:
\begin{align}
 i p^\mu \G^{(1;1,0)}_{A_\nu(-p) A_\mu(p)} + ev \G^{(1;1,0)}_{A_\nu(-p) \chi(p)}  = 0 \, , 
\end{align}
in agreement with Eq.(\ref{sti.decomp}).
%
%%%%%
%

\begin{figure}[h]
\centering
\begin{tabular}{cc}
\begin{minipage}{.5\textwidth}
  \centering
 \begin{tikzpicture}
     \begin{feynman}
        \vertex (a1);
        \vertex[right=1.5cm of a1] (a2);
        \vertex[right=1.5cm of a2] (f1);
        \vertex[above=1cm of f1] (a3);
        \vertex[right=1.5cm of f1] (a4);         
        \vertex[right=1.5cm of a4] (a5);
        \diagram* {
            (a1) -- [photon] (a2)  -- [photon, half left, looseness=1.5, color=red] (a4) -- [photon] (a5),
            (a2) --  (a4) 
         };
    \end{feynman}
 \end{tikzpicture}
\end{minipage}
&
\begin{minipage}{.5\textwidth}
 \centering
 \begin{tikzpicture}
    \begin{feynman}
        \vertex (a1);
        \vertex[right=1.5cm of a1] (a2);
        \vertex[right=1.5cm of a2] (f1);
        \vertex[above=1cm of f1] (a3);
        \vertex[right=1.5cm of f1] (a4);         
        \vertex[right=1.5cm of a4] (a5);
        \diagram* {
            (a1) -- [photon] (a2)  -- [photon, half left, looseness=0.8, color=red] (a3) -- [photon, half left, looseness=0.8] (a4) -- [photon] (a5),
            (a2) -- (a4) 
         };
    \end{feynman}
   \end{tikzpicture}
\end{minipage}
\cr
\begin{minipage}{.5\textwidth}
  \centering
  $(\alpha)$
\end{minipage}
&
\begin{minipage}{.5\textwidth}
  \centering
  $(\beta)$
\end{minipage}
\end{tabular}
\caption{
{\small 
Contributions to the 1-PI amplitude $\G^{(1)}_{A_\mu A_\nu}$ from the $X_\mu$-dependent interaction vertices.
}
}
\label{figure:3}
\end{figure}
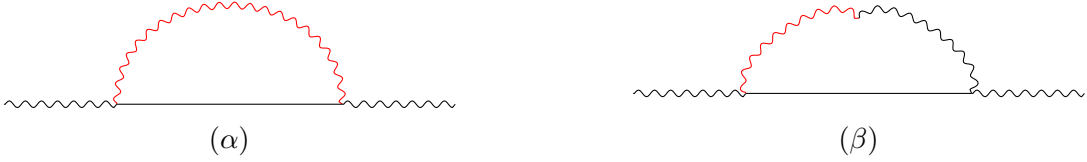
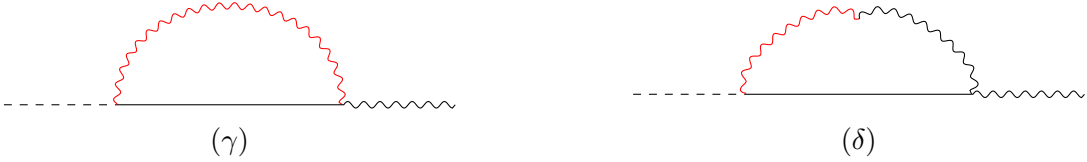
\begin{figure}[h]
\centering
\begin{tabular}{cc}
\begin{minipage}{.5\textwidth}
  \centering
 \begin{tikzpicture}
     \begin{feynman}
        \vertex (a1);
        \vertex[right=1.5cm of a1] (a2);
        \vertex[right=1.5cm of a2] (f1);
        \vertex[above=1cm of f1] (a3);
        \vertex[right=1.5cm of f1] (a4);         
        \vertex[right=1.5cm of a4] (a5);
        \diagram* {
            (a1) -- [scalar] (a2)  -- [photon, half left, looseness=1.5, color=red] (a4) -- [photon] (a5),
            (a2) --  (a4) 
         };
    \end{feynman}
 \end{tikzpicture}
\end{minipage}
&
\begin{minipage}{.5\textwidth}
 \centering
 \begin{tikzpicture}
    \begin{feynman}
        \vertex (a1);
        \vertex[right=1.5cm of a1] (a2);
        \vertex[right=1.5cm of a2] (f1);
        \vertex[above=1cm of f1] (a3);
        \vertex[right=1.5cm of f1] (a4);         
        \vertex[right=1.5cm of a4] (a5);
        \diagram* {
            (a1) -- [scalar] (a2)  -- [photon, half left, looseness=0.8, color=red] (a3) -- [photon, half left, looseness=0.8] (a4) -- [photon] (a5),
            (a2) -- (a4) 
         };
    \end{feynman}
   \end{tikzpicture}
\end{minipage}
\cr
\begin{minipage}{.5\textwidth}
  \centering
  $(\gamma)$
\end{minipage}
&
\begin{minipage}{.5\textwidth}
  \centering
  $(\delta)$
\end{minipage}
\end{tabular}
\caption{
{\small 
Contributions to the 1-PI amplitude $\G^{(1)}_{\chi A_\nu}$ from the $X_\mu$-dependent interaction vertices.
}
}
\label{figure:4}
\end{figure}

\item $(1,1)$-sector

The analysis proceeds in a similar way. One needs to keep only those diagrams with one internal $h$- and one $a'_\mu$-line.
Therefore one gets contributions from diagram (i) in Fig.~\ref{figure:1}, (I) in Fig.~\ref{figure:2} and from all diagrams
in Fig.~\ref{figure:3} and \ref{figure:4}. 
Their expressions are reported in 
Eqs.(\ref{symm.contr}) and (\ref{xmu.contr}).
Again by summing all diagrams one can check by direct computation that the
ST identity holds when projected in the $(1,1)$-sector:
\begin{align}
 i p^\mu \G^{(1;1,1)}_{A_\nu(-p) A_\mu(p)} + ev \G^{(1;1,1)}_{A_\nu(-p) \chi(p)}  = 0 \, . 
\end{align}
We notice that by Eq.(\ref{prop.XX.XA}) all diagrams with an internal $\Delta_{X_\mu X_\nu}$ or $\Delta_{X_\mu A_\nu}$ propagator
have a $(1,1)$-projection that is the opposite of the $(1,0)$-projection, since $\Delta_{X''_\mu X''_\nu}=-\Delta_{a'_\mu a'_\nu}$.
Hence when summing over all projections in order to recover the full amplitudes they will cancel out against each other.

The second identity in Eqs.~(\ref{sti.2pt.1}) can be verified along 
similar lines. We notice that by combining
the first and the second
of Eqs.(\ref{sti.2pt.1}) one obtains an identity that fixes uniquely 
in a purely algebraic way
the two-point function of the pseudo-Goldstone fields:
\begin{align}
\G^{(1)}_{\chi(-p)\chi(p)} =
\frac{p^\mu p^\nu}{ev}  
\G^{(1)}_{A_\mu(-p) A_\nu(p)} \, .
\label{sti.chichi}
\end{align}

This is a general feature of spontaneously broken gauge theories.
In fact, in the sector with ghost number zero amplitudes with at least one pseudo-Goldstone leg are fixed by amplitudes without by the ST identities. Also Eq.(\ref{sti.chichi})
can be projected in the sectors
$(\l,\l')$.

The third identity is trivially verified since the field $a_\mu$ does not interact in the symmetric basis, as a consequence of the first of Eq.(\ref{ha.eqs.j}).

\end{itemize}
\section{Flows in the gauge-invariant kinetic and mass terms}\label{sec.pde}

In addition to being an effective tool in
order to derive the decomposition of the 
ST identities in separately invariant sectors,
the kinetic and mass terms in the 
gauge-invariant variables yield
non power-counting renormalizable
terms that are equipped with
some novel differential equations. The latter
strongly constrain the dependence
of 1-PI amplitudes on such parameters.

%{\tt Revised until here ... January 17, 2024}

First of all we notice that the mass and kinetic terms in Eqs.(\ref{scalar.mass}),
(\ref{new.mass}),
(\ref{new.kin.term.landau}) 
and (\ref{new.kin.scalar}) are the only
allowed deformations with at most two derivatives  preserving the $h$ and 
$a_\mu$-equations at the quantum level.
I.e. they modify the
equations according to
\begin{align}
& \frac{\delta \G^{(0)}}{\delta h} = (\square + m^2)X + \bar c^* - (z \square h  + M^2 - m^2) h \, , 
\end{align}
\begin{align}
& \frac{\delta \G^{(0)}}{\delta a_\mu} = \Sigma^{\mu\nu}_{(\xi)} X_v + \bar c^*_\mu + z' (\square g^{\mu \nu} - \partial^\mu \partial^\nu) a'_\nu + M_a^2 a'_\mu
\end{align}
and, since the r.h.s. of the above equations are linear in the quantum fields, no further external source is needed to control their renormalization.
Consequently
\begin{align}
     \frac{\delta \G^{(j)}}{\delta h} = 0 \, , \qquad
    \frac{\delta \G^{(j)}}{\delta a_\mu} = 0 \, ,  \qquad 
    j \geq 1 \, .
\end{align}

The parameters $z, M^2$ and $z', \Mp2$ only enter in the  propagators 
$\Delta_{hh}$ and $\Delta_{a'_{\mu} a'_{\nu}}$ respectively and never in the interaction vertices.
Moreover the propagator $\Delta_{hh}$ is an eigenvector of
eigenvalue $-1$ of the differential operator
\begin{align}
    \dEulS \equiv (1+z) \frac{\partial}{\partial z} + M^2 \frac{\partial}{\partial M^2} \, ,
\end{align}
while 
the propagator $\Delta_{a'_{\mu} a'_{\nu}}$ is an eigenvector of eigenvalue $-1$ of the differential operator
\begin{align}
    \dEulV \equiv (1+z') \frac{\partial}{\partial z'} + \Mp2 \frac{\partial}{\partial \Mp2} \, ,
\end{align}
i.e.
\begin{align}
    \dEulS \Delta_{hh} = - \Delta_{hh} \, , \qquad
    \dEulV \Delta_{a'_{\mu} a'_{\nu}} = - \Delta_{a'_{\mu} a'_{\nu}} \, ,
\end{align}
as can be easily seen by direct computation.
We now consider a $n$-th loop 1-PI amplitude $\G^{(n)}_{\varphi_1 \dots \varphi_r}$ with $r$ $\varphi_i$
external legs, $\varphi_i = \varphi(p_i)$ denoting
a generic field or external source of the theory with incoming momentum $p_i$.

$\G^{(n)}_{\varphi_1 \dots \varphi_r}$ can again be decomposed as the sum
of all diagrams with (amputated) external legs $\varphi_1 \dots \varphi_r$ with  zero, one, two, $\dots$ $\lS$ internal
$h$-propagators:
and
zero, one, two, $\dots$ $\lV$ internal
$a'$-propagators:
\begin{align}
    \G^{(n)}_{\varphi_1 \dots \varphi_r} = \sum_{\lS,\lV \geq 0}
    \G^{(n; \lS, \lV)}_{\varphi_1 \dots \varphi_r}
\end{align}

Each $\G^{(n; \lS, \lV)}_{\varphi_1 \dots \varphi_r}$ is an eigenvector of $\dEulS$ of eigenvalue $-\lS$  and of
$\dEulV$ of eigenvalue
$-\lV$,
namely
\begin{align}
    \dEulS \G^{(n;  \lS, \lV)}_{\varphi_1 \dots \varphi_r} = 
    - \lS \G^{(n; \lS, \lV)}_{\varphi_1 \dots \varphi_r}\, , 
    \qquad
    \dEulV \G^{(n;  \lS, \lV)}_{\varphi_1 \dots \varphi_r} = 
    - \lV \G^{(n; \lS, \lV)}_{\varphi_1 \dots \varphi_r}\, .  
    \label{pdeq.euler}
\end{align}
The above partial differential equations can be solved in terms of homogeneous function in the variables $M^2/(1+z)$ and $\Mp2/(1+z')$.
I.e. the most general solution 
to Eqs.(\ref{pdeq.euler})
can be written as 
\begin{align}
\G^{(n; \lS, \lV)}_{\varphi_1 \dots \varphi_r} (z, M^2, z', \Mp2) = 
 \frac{1}{(1+z)^{\lS}} \frac{1}{(1+z')^{\lV}}\G^{(n; \lS, \lV)}_{\varphi_1 \dots \varphi_r} \Big ( 0, \frac{M^2}{1+z}, 0, \frac{\Mp2}{1+z'} \Big ) \, .
 \label{scaling}
\end{align}
This result holds true to all orders in the loop expansion
and is valid for the fully renormalized amplitudes, provided that the finite normalization conditions are chosen in such a way to fulfill Eq.(\ref{scaling})~\cite{Binosi:2022ycu}.

We observe that 
Eq.(\ref{scaling}) predicts the full dependence on the coefficients of the gauge-invariant mass and kinetic terms, i.e. it yields the exact dependence on $z,z',M^2, \Mp2$.

In this sense it provides a tool for resummation at fixed order in the loop expansion.

This is a non-trivial application of the 
gauge-invariant field formalism.

Let us see how the computation would proceed in the conventional approach.
One would start from the action of the Abelian Higgs-Kibble model with fermions and supplement it by the higher-dimensional operators
\begin{align}
\int \d \, \Big \{ &
    -\frac{2 z'}{e^2 v^4} 
    \phi^\dagger D_\mu \phi 
    (\square g^{\mu\nu} - \partial^\mu \partial^\nu )  \phi^\dagger D_\nu \phi
    \nonumber \\
& + \frac{\mu^2 - M_A^2}{2 v^2 M_A^2}  
\phi^\dagger \phi \Big [ 4 (D^\mu \phi)^\dagger D_\mu \phi +
2 \partial^\mu ( \phi^\dagger D_\mu \phi + (D_\mu \phi)^\dagger \phi ) - \square \phi^\dagger \phi
\Big ] 
%\nonumber \\
%& 
- \frac{z}{2v^2} \phi^\dagger \phi \square \phi^\dagger \phi \Big \}
\end{align}
that are obtained 
from the operators in Eqs.(\ref{non.ren.term}),
(\ref{new.kin.term.landau}) and 
(\ref{new.kin.scalar}) (we drop the cohomologially trivial $b$-dependent terms in Eq.(\ref{mass.term.prime}) as
explained in Eq.(\ref{mass.cohom.triv}))
by going on-shell with the $X$ and $X_\mu$-fields. 

The resulting theory contains a huge number of 
dim.6 and dim.8 interaction vertices in $z,z',\mu^2$ and the problem is to resum the exact dependence of the 1-PI amplitudes in $z,z',\mu^2$
at any given order in the loop expansion.
To say the least, this turns out to be an involved computational task.

On the other hand, in the gauge-invariant field formalism resummation is easily established by the flow equation and compactly stated in Eq.(\ref{scaling}).

The significant computational advantage in dealing with these operators should be immediately clear.

We stress that Eq.(\ref{scaling}) holds true to all orders in perturbation theory.
Phenomenological implications have to be studied.
As an example, it implies  non-trivial decoupling limits in Higgs portal theories, as discussed in Ref.~\cite{Quadri:2024xuv}.

\section{Conclusions}\label{sec.concl}

In the present paper  gauge-invariant variables for
scalar and gauge fields in a spontaneously broken theory have been constructed.
While this is always possible for scalars and vectors,
particular conditions on the fermion charges must be met
in order to introduce gauge-invariant variables for fermionic fields.
These conditions are fulfilled in the physically relevant case of the SM.

Gauge-invariant variables provide additional information w.r.t. the conventional formalism
on the decomposition of the ST identities into subsets of
separately invariant diagrams. 

Moreover the functional identities controlling the dynamics
of such fields are compatible with a specific set of  kinetic
and mass operators for the scalar and the vector field.

These operators can be introduced in order to obtain an elegant algebraic proof
of the ST decomposition onto separately invariant sectors.

On top of that, being quadratic in the gauge-invariant fields,
they only affect propagators.

This simple fact allows to write
a set of powerful partial differential equations resulting
in the prediction of the exact dependence of the 1-PI
amplitudes on the parameters associated with those
operators.

It should be noticed that once these operators are treated in the 
conventional formalism they give rise to complicated
non-renormalizable interactions, that make it hard to
guess the resummation properties easily established 
in the gauge-invariant field approach.

This does not come as a surprise, since gauge-invariant variables can be thought of as the field representation most suited to the description of gauge-invariant fields.

The results obtained in the present paper generalize those already established for the scalar case.
They pave the way for further work, in particular the formulation of the SM in the gauge-invariant fields formalism and a detailed analysis of the BSM models in the presence of the quadratic mass and kinetic terms in the gauge-invariant fields, as well as the associated resummations at fixed order in the loop expansion.

\appendix

\section*{Acknowledgments}

Warmful hospitality at the Mainz Institute for Theoretical Physics (MITP) during the MITP Scientific Program ``EFT Foundations and Tools" in August 2023, where part of this work has been carried out, is gratefully acknowledged.
Useful discussions with D.~Stoeckinger and S.~Dittmaier are also acknowledged.

\section{The model and its symmetries}\label{app.cl.act}

We start from an Abelian Higgs-Kibble model with chiral fermions and the usual quartic Higgs potential:
\begin{align}
    S = \int \d \, \Big [ 
    - \frac{1}{4} F_{\mu\nu}^2 + (D^\mu \phi)^\dagger D_\mu \phi -
    \frac{m^2}{2v^2} \Big ( \phi^\dagger \phi - \frac{v^2}{2} \Big )^2 + i \bpsi \Ds \psi + \frac{G}{\sqrt{2}} \bpsi ( 1 - \gamma_5) \psi \phi + {\mbox{h.c.}}
    \Big ]
\label{cl.act}
\end{align}
$\phi$ is the Higgs field
\begin{align}
\phi = \frac{1}{\sqrt{2}} ( v + \sigma + i \chi ) \, ,
\end{align}
$\sigma$ describing the physical Higgs mode and $\chi$ the pseudo-Goldstone.
The covariant derivative on the scalar $\phi$
is defined as
\begin{align}
D_\mu \phi = \partial_\mu \phi - i e A_\mu \phi \, ,
\end{align}
while on the fermion field it acts as
\begin{align}
D_\mu \psi = \partial_\mu \psi - i \frac{e}{2} \gamma_5 \psi \, .
\end{align}
The mass of the gauge field $A_\mu$ is $M_A = ev$, the mass of the fermion field is $m_e=Gv$.

The gauge-fixing and ghost terms in Landau gauge
are given by
\begin{align}
    S_{\tiny{\mbox{g.f. + ghost}}} = 
    \int \d \Big [ - b \partial A + \bar \omega \square \omega \Big ] \, ,
\end{align}
$b$ being the Nakanishi-Lautrup field and $\bar \omega, \omega$
the antighost and ghost fields.

In a generic $R_\xi$-gauge the gauge-fixing
and the associated ghost terms read
\begin{align}
    S_{\tiny{\mbox{g.f. + ghost}, \xi}} = 
    \int \d \Big [ \frac{\xi}{2} b^2 - b \Big ( \partial A + \xi M_A \chi \Big ) + \bar \omega \square \omega + \xi e^2 v \bar \omega \omega ( v + \sigma )  \Big ] \, .
    \label{rxi.gauge}
\end{align}
The Landau gauge is recovered in the limit
$\xi\rightarrow 0$.

In all gauges the gauge-fixed action is invariant
under the following BRST symmetry:
\begin{align}
    & s A_\mu = \partial_\mu \omega \, , \qquad 
    s \sigma = - e \omega \chi \, , \qquad
    s \chi = e \omega (v + \sigma) \, , \qquad
    s \phi = i e \omega \phi \, , \nonumber \\
    & s \psi = -i \frac{e}{2} \gamma_5 \psi \omega \, ,
    \qquad s \bpsi = i \frac{e}{2} \omega \bpsi \gamma_5 \, , \qquad s \bar \omega = b \, , \qquad s b =0 \, ,
    \qquad s \omega = 0 \, .
    \label{gauge.brst}
\end{align}

%%%%%%%%%
Let us now derive the complete vertex
functional in the presence of the
extended set of gauge-invariant dynamical fields.
For that purpose one needs to add the
terms implementing the constraints
in the scalar and gauge sectors in 
Eqs.(\ref{aux.scalar}) and (\ref{X.term}).
Moreover one also needs to add the
antifields associated
to the $\s$-variations of $\bar c^*, \bar c^*_\mu$ (the only $\s$-variations that
are non-linear in the quantum fields and
thus undergo an independent renormalization
to be controlled by the relevant 
antifields).

The full classical vertex functional is 
\begin{align}
    \G^{(0)} = \int \d  \Big \{ &
        - \frac{1}{4} F_{\mu\nu}^2 + (D^\mu \phi)^\dagger D_\mu \phi -
        \frac{m^2}{2v^2} \Big ( \phi^\dagger \phi - \frac{v^2}{2} \Big )^2 + i \bpsi \Ds \psi + \frac{G}{\sqrt{2}} \bpsi ( 1 - \gamma_5) \psi \phi + {\mbox{h.c.}} \nonumber \\
        & + \frac{\xi}{2} b^2 - b \Big ( \partial A + \xi M_A \chi \Big ) + \bar \omega \square \omega + \xi e^2 v \bar \omega \omega ( v + \sigma ) \nonumber \\
        & + X (\square + m^2) \Big [ h -  \frac{1}{v} \Big ( \phi^\dagger \phi - \frac{v^2}{2} \Big ) \Big ] 
        - \frac{M^2 - m^2}{2} h^2
        - \bar c (\square + m^2) c \nonumber \\
        & - \bar c_\mu \Sigma_{(\xi)}^{\mu\nu} c_\nu +
        X_\mu \Sigma_{(\xi)}^{\mu\nu} 
        \Big [ a_\nu - \frac{i}{ev^2} \Big ( 2 \phi^\dagger  D_\nu \phi - \partial_\nu ( \phi^\dagger \phi ) \Big ) \Big ] \nonumber \\
        & - \sigma^* e \omega \chi + \chi^* e \omega ( v + \sigma) - \frac{ i e }{2} \bar \eta  \gamma_5 \psi \omega + \frac{ i e }{2} \omega \bar \psi \gamma_5 \eta \nonumber \\
        & + \bar c^* \Big [ h -  \frac{1}{v} \Big ( \phi^\dagger \phi - \frac{v^2}{2} \Big ) \Big ]
        + \bar c^*_\mu \Big [ a^\mu - \frac{i}{ev^2} \Big ( 2 \phi^\dagger  D^\mu \phi - \partial^\mu ( \phi^\dagger \phi ) \Big )\Big ]
    \Big \} \, .
    \label{cl.vertex.functional}
\end{align}
The operator $\Sigma_{(\xi)}^{\mu\nu}$ 
in Eq.(\ref{cl.vertex.functional}) is 
given by 
\begin{align}
\Sigma_{(\xi)}^{\mu\nu} \equiv \Big [ \square g^{\mu\nu} - \Big ( 1 - \frac{1}{\xi} \Big ) \partial^\mu \partial^\nu \Big ] + M_A^2 g^{\mu\nu} 
\end{align}
for $\xi \neq 0$ and by
\begin{align}
    \Sigma_{(0)}^{\mu\nu} \equiv (\square g^{\mu\nu} - \partial^\mu \partial^\nu) + M_A^2 g^{\mu\nu}
\end{align}
for the $\xi=0$ case.

\section{Functional Identities}\label{app.funct.ids}

The vertex functional $\G^{(0)}$ obeys several identities, that can be summarized as follows:
\begin{itemize}
\item the $b$-equation
\begin{align}
    \frac{\delta \G^{(0)}}{\delta b} = \xi b - \partial A - \xi M_A \chi \, ;
    \label{b.eq}
\end{align}
\item the ghost equation
\begin{align}
    \frac{\delta \G^{(0)}}{\delta \bar \omega} = \square \omega + \xi M_A \frac{\delta \G^{(0)}}{\delta \chi^*} \, ;
    \label{gh.eq}
\end{align}
\item 
the Slavnov-Taylor (ST) identity associated with the U(1) gauge group:
\begin{align}
    {\cal S} (\G^{(0)}) =
    \int \d \, \Big [ & \partial_\mu \omega 
    \frac{\delta \G^{(0)}}{\delta A_\mu} + 
    \frac{\delta \G^{(0)}}{\delta \sigma^*}
    \frac{\delta \G^{(0)}}{\delta \sigma} +
    \frac{\delta \G^{(0)}}{\delta \chi^*}
    \frac{\delta \G^{(0)}}{\delta \chi} \nonumber \\
    & +
    \frac{\delta \G^{(0)}}{\delta \bar \eta}
    \frac{\delta \G^{(0)}}{\delta \psi} +
    \frac{\delta \G^{(0)}}{\delta \eta}
    \frac{\delta \G^{(0)}}{\delta \bar \psi} 
    + b \frac{\delta \G^{(0)}}{\delta \bar \omega}
     \Big ] = 0 \, ;
     \label{sti}
\end{align}
\item 
the $X$-equation
\begin{align}
    \frac{\delta \G^{(0)}}{\delta X} = 
    (\square + m^2) \frac{\delta \G^{(0)}}{\delta \bar c^*} ;
    \label{X.eq}
\end{align}
\item 
the $X_\mu$-equation
\begin{align}
    \frac{\delta \G^{(0)}}{\delta X_\mu} =
    \Sigma_{(\xi)}^{\mu\nu} \frac{\delta \G^{(0)}}{\delta \bar c^{\nu *}} \, ;
    \label{Xmu.eq}
\end{align}
\item the ghost-antighost equations for the 
constraint scalar and vector ghost and antighost fields
\begin{align}
    \frac{\delta \G^{(0)}}{\delta \bar c } = 
    - (\square + m^2) c \, , \qquad
    \frac{\delta \G^{(0)}}{\delta  c } = 
    (\square + m^2) \bar c  \, ;
    \label{barc.c.eq}
\end{align}
\begin{align}
    \frac{\delta \G^{(0)}}{\delta \bar c^\mu } = - \Sigma_{(\xi)}^{\mu\nu} 
    c_\nu  \, , \qquad
    \frac{\delta \G^{(0)}}{\delta  c^\mu } = 
    \Sigma_{(\xi)}^{\mu\nu} \bar c_\nu  \, ;
    \label{vect.barc.c.eq}
\end{align}
\item the scalar and vector constraint ST identities
\begin{align}
    & {\cal S}_{\tiny \s, \mbox{scal}} (\G^{(0)}) = 
    \int \d \, \Big [
    c \frac{\delta \G^{(0)}}{\delta X} + \frac{\delta \G^{(0)}}{\delta \bar c^*}
    \frac{\delta \G^{(0)}}{\delta \bar c} 
    \Big ] = 0 \, , 
    \label{scal.constr.sti}
    \\
    & {\cal S}_{\tiny \s, \mbox{vect}}  (\G^{(0)}) = 
    \int \d \, \Big [
    c_\mu \frac{\delta \G^{(0)}}{\delta X_\mu} + \frac{\delta \G^{(0)}}{\delta \bar c_\mu^*}
    \frac{\delta \G^{(0)}}{\delta \bar c^\mu} 
    \Big ] = 0 \, . 
    \label{vect.constr.sti}
\end{align}
The constraint ST identities in Eqs.(\ref{scal.constr.sti}) and (\ref{vect.constr.sti})
are equivalent to the $X$- and 
$X_\mu$-equations (\ref{X.eq}) and (\ref{Xmu.eq}) since the
constraint ghost and antighost fields are free
by Eqs.(\ref{barc.c.eq}) and (\ref{vect.barc.c.eq}).
\item the $h$- and $a_\mu$-equations:
\begin{align}
    \frac{\delta \G^{(0)}}{\delta h} = (\square + m^2) X - (M^2 - m^2 ) h + \bar c^* \, , 
    \label{h.eq}
\end{align}
and
\begin{align}
    \frac{\delta \G^{(0)}}{\delta a_\mu} =     \Sigma_{(\xi)}^{\mu\nu} X_\nu + \bar c^{*\mu} \, . 
    \label{amu.eq}   
\end{align}

\end{itemize}

\section{Scalar Sector}\label{app:sector}

Diagonalization of the quadratic part
in the sector spanned by $X, \sigma$ and $h$
is achieved by setting
\begin{align}
    X = -X' - h \, , \qquad \sigma = \sigma' + X' + h \, .
    \label{scalar.fred}
\end{align}
The propagators in the mass eigenstates basis
are
\begin{align}
    \Delta_{\sigma'\sigma'} = - \Delta_{X'X'} = 
    \frac{i}{p^2 - m^2} \, , \qquad
    \Delta_{hh} = \frac{i}{p^2 - M^2} \, .
\end{align}
In the symmetric basis the propagators
are
\begin{align}
    \Delta_{XX} = \Delta_{X\sigma} = - \frac{i (M^2 - m^2)}{(p^2 - M^2) (p^2 - m^2)} \, , \qquad
    \Delta_{hh} = \Delta_{\sigma h} = \Delta_{\sigma \sigma} = - \Delta_{X h} = \frac{i}{p^2 - M^2} \, .
    \label{scalar.symm.props}
\end{align}

\section{Landau Gauge}\label{app:landau}

One must diagonalize the quadratic part given by 
\begin{align}
     \int \d \Big \{ & 
    \frac{1}{2} A_\mu ( \square g^{\mu\nu} - \partial^\mu \partial^\nu ) A_\nu + \frac{M_A^2}{2} \Big ( A_\mu - \frac{1}{M_A} \partial_\mu \chi \Big )^2 - b \partial A \nonumber \\ 
    & + X_\mu \Big [ (\square g^{\mu\nu} - \partial^\mu \partial^\nu) + M_A^2 g^{\mu\nu}  \Big ] \Big ( a_\nu - A_\nu + \frac{1}{M_A} \partial_\nu \chi \Big ) \Big \} \, .
    \label{quad.landau.0}
\end{align}
One first removes the $\chi-A_\mu$-mixing via the redefinition
\begin{align}
    b = b' + M_A \chi \, ,
    \label{loc.field.redef.0}
\end{align}
followed by the cancellation of the $b'-A_\mu$-mixing by the replacement
\begin{align}
A_\mu = A'_\mu - \frac{1}{M_A^2}\partial_\mu b' \, .
\end{align}
A further set of field redefinitions
\begin{align}
    A'_\mu = A''_\mu + X_\mu \, , \qquad X_\mu = X'_\mu + a_\mu 
\end{align}
takes care of the $X_\mu-A_\nu$ and $X_\mu-a_\nu$ mixing. 
One is eventually left with
\begin{align}
     \int \d \Big \{ & 
    \frac{1}{2} A''_\mu [ (\square + M_A^2) g^{\mu\nu} - \partial^\mu \partial^\nu ] A''_\nu 
    -
    \frac{1}{2} X'_\mu [ (\square + M_A^2) g^{\mu\nu} - \partial^\mu \partial^\nu ] X'_\nu
    \nonumber \\
    & +
     \frac{1}{2} a_\mu [ (\square + M_A^2) g^{\mu\nu} - \partial^\mu \partial^\nu ] a_\nu 
  - \frac{1}{2 M_A^2} \partial^\mu b'\partial_\mu b'
    + \frac{1}{2} \partial^\mu \chi \partial_\mu \chi   \nonumber \\ 
  &
  + ( X'_\mu + a_\mu ) \partial^\mu (b' + M_A \chi ) \Big \} \, .  
    \label{quad.landau.1}
\end{align}
The mixing terms in the last line of the above equation can be removed by the local field redefinition 
\begin{align}
    X'_\mu = X''_\mu + \frac{1}{M_A^2} \partial_\mu b' + \frac{1}{M_A} \partial_\mu \chi \, , \qquad
    a_\mu = a'_\mu -\frac{1}{M_A^2} \partial_\mu b' - \frac{1}{M_A} \partial_\mu \chi \, .
    \label{loc.field.redef.2}
\end{align}
No new $b'-\chi$-mixing is generated. In momentum space the diagonal propagators  (mass eigenstates) are finally given by
\begin{align}
    \Delta_{A''_\mu A''_\nu} = \Delta_{a'_\mu a'_\nu} = - \Delta_{X''_\mu X''_\nu} = \frac{i}{-p^2 + M_A^2} T_{\mu\nu} + \frac{i}{M_A^2}L_{\mu\nu} \, , 
    ~~
    \Delta_{b'b'} = -\frac{i M_A^2}{p^2} \, , ~~
    \Delta_{\chi\chi} = \frac{i}{p^2} \, .
    \label{landau.gauge.props}
\end{align}
In the symmetric basis $(A_\mu, \chi, b, X_\mu, a_\mu)$ 
\begin{align}
& b = b' + M_A \chi \, , \qquad A_\mu = A''_\mu+X''_\mu+a'_\mu - \frac{1}{M_A^2}\partial_\mu b'\, , \nonumber \\
& X_\mu = X''_\mu + a'_\mu \, , \qquad
a_\mu = a'_\mu - \frac{1}{M_A^2}\partial_\mu b' - \frac{1}{M_A} \partial_\mu \chi 
\label{landau.diag}
\end{align}
the non-vanishing propagators
are given by
\begin{align}
    &
    \Delta_{A\mu A_\nu} = \frac{i}{-p^2 + M_A^2} T_{\mu\nu} \, , \qquad
    \Delta_{A_\mu b} = - \frac{p_\mu}{p^2}\, , \qquad
    \Delta_{A_\mu a_\nu} = \frac{i}{-p^2 + M_A^2} T_{\mu\nu} \, , \nonumber \\
    & \Delta_{X_\mu a_\nu} = \frac{i}{-p^2 + M_A^2} T_{\mu\nu} + \frac{i}{M_A^2} L_{\mu\nu} \, , \qquad
    \Delta_{a_\mu a_\nu} = \frac{i}{-p^2 + M_A^2} T_{\mu\nu} + \frac{i}{M_A^2} L_{\mu\nu} \, , \nonumber \\
    & \Delta_{a_\mu \chi} = \frac{1}{M_A}\frac{p_\mu}{p^2} \, , \qquad
    \Delta_{b \chi} = \frac{i M_A}{p^2} \, , \qquad
    \Delta_{\chi\chi} = \frac{i}{p^2} \, .
    \label{landau.prop.symm}
\end{align}

The mass eigenstates are expressed in terms of the symmetric basis fields as
\begin{align}
    & b' = b - M_A \chi \, , \qquad
       A''_\mu = A_\mu - X_\mu + \frac{1}{M_A^2}\partial_\mu b - \frac{1}{M_A}\partial_\mu \chi \, , \nonumber\\ 
    & X''_\mu = X_\mu - a_\mu - \frac{1}{M_A^2} \partial_\mu b \, , \qquad 
    a'_\mu = a_\mu + \frac{1}{M_A^2} \partial_\mu b \, .
    \label{landau.vec.fred}
\end{align}

\section{$R_\xi$-gauge}\label{app:rxi}
By setting
$$ b' = b - \frac{1}{\xi} (\partial A + \xi M_A \chi)$$
the relevant quadratic terms are given by
\begin{align}
    \int \d \Big \{ & 
    \frac{1}{2} A_\mu ( \square g^{\mu\nu} - \partial^\mu \partial^\nu ) A_\nu + \frac{M_A^2}{2} A^2_\mu 
    -\frac{1}{2\xi} (\partial A)^2 +
    \frac{1}{2} \partial^\mu \chi \partial_\mu \chi -
    \frac{\xi}{2} M_A^2 \chi^2
    + \frac{\xi}{2} {b'}^2
    \nonumber \\ 
    & + X_\mu \Big [ (\square g^{\mu\nu} - \Big (1 - \frac{1}{\xi} \Big ) \partial^\mu \partial^\nu)  + M_A^2  g^{\mu\nu}  \Big ] \Big ( a_\nu - A_\nu + \frac{1}{M_A} \partial_\nu \chi \Big ) \Big \} \, .
\end{align}
Diagonalization is obtained by first replacing
$A_\mu = A'_\mu + X_\mu$ to remove
the $A-X$-mixing and then by redefining
$X_\mu = X'_\mu + a_\mu$ in order to cancel the 
transverse part of the $X-a$-mixing.
One is left with
\begin{align}
    \int \d \Big \{ &
    \frac{1}{2} A'_\mu 
    \Big [ (\square g^{\mu\nu} - \Big (1 - \frac{1}{\xi} \Big ) \partial^\mu \partial^\nu)  + M_A^2  g^{\mu\nu}  \Big ]  A'_\nu 
    - \frac{1}{2} X'_\mu \Big [ (\square g^{\mu\nu} - \Big (1 - \frac{1}{\xi} \Big ) \partial^\mu \partial^\nu)  + M_A^2  g^{\mu\nu}  \Big ] X'_\nu \nonumber \\
    & + \frac{1}{2} a_\mu \Big [ (\square g^{\mu\nu} - \Big (1 - \frac{1}{\xi} \Big ) \partial^\mu \partial^\nu)  + M_A^2  g^{\mu\nu}  \Big ] a_\nu - \frac{1}{2} \chi ( \square + \xi M_A^2) \chi \nonumber \\
    & + \frac{1}{\xi M_A} ( X'_\mu + a_{\mu}  ) 
    ( \square + \xi M_A^2  ) \partial^\mu \chi
   \Big \} \, .
\end{align}
In order to remove the mixing in the last term of the above equation we redefine
\begin{align}
    X'_\mu = X''_\mu + \frac{1}{M_A} \partial_\mu \chi \, , \qquad
    a_\mu = a'_\mu - \frac{1}{M_A}  \, ,
    \partial_\mu \chi
    \label{rxi.redef}
\end{align}
yielding finally
\begin{align}
    \int \d \Big \{ &
    \frac{1}{2} A'_\mu 
    \Big [ (\square g^{\mu\nu} - \Big (1 - \frac{1}{\xi} \Big ) \partial^\mu \partial^\nu)  + M_A^2  g^{\mu\nu}  \Big ]  A'_\nu 
    - \frac{1}{2} X''_\mu \Big [ (\square g^{\mu\nu} - \Big (1 - \frac{1}{\xi} \Big ) \partial^\mu \partial^\nu)  + M_A^2  g^{\mu\nu}  \Big ] X''_\nu \nonumber \\
    & + \frac{1}{2} a'_\mu \Big [ (\square g^{\mu\nu} - \Big (1 - \frac{1}{\xi} \Big ) \partial^\mu \partial^\nu)  + M_A^2  g^{\mu\nu}  \Big ] a'_\nu - \frac{1}{2} \chi ( \square + \xi M_A^2) \chi + \frac{\xi}{2} {b'}^2
   \Big \} \, .
   \label{rxi.1}
\end{align}

In the diagonal basis the propagators are
\begin{align}
    \Delta_{A'_\mu A'_\nu} = \Delta_{a'_\mu a'_\nu} = - \Delta_{X''_\mu X''_\nu} = \frac{i}{-p^2 + M_A^2} T_{\mu\nu} - \frac{i \xi}{p^2 - \xi M_A^2}L_{\mu\nu} \, , 
    ~~
    \Delta_{b'b'} = \frac{i}{\xi} \, , ~~
    \Delta_{\chi\chi} = \frac{i}{p^2 - \xi M_A^2} \, .
    \label{rxi.diag}
\end{align}

In the symmetric basis $(A_\mu, \chi, b, X_\mu, a_\mu)$ 
\begin{align}
& b = b' + \frac{1}{\xi} ( \partial A' + \partial X'' + \partial a') + M_A \chi \, , \qquad A_\mu = A'_\mu+X''_\mu+a'_\mu \, , \nonumber \\
& X_\mu = X''_\mu + a'_\mu \, , \qquad
a_\mu = a'_\mu  - \frac{1}{M_A} \partial_\mu \chi 
 \label{rxi.symm}
\end{align}
the non-vanishing propagators
are given by
\begin{align}
    &
    \Delta_{A\mu A_\nu} =  \Delta_{A_\mu a_\nu} = \frac{i}{-p^2 + M_A^2} T_{\mu\nu} - \frac{i \xi}{p^2 - \xi M_A^2} L_{\mu\nu} \, , \qquad
    \Delta_{A_\mu b} = - \frac{p_\mu}{p^2 - \xi M_A^2}\, , 
    \nonumber \\
    & \Delta_{X_\mu a_\nu} = \frac{i}{-p^2 + M_A^2} T_{\mu\nu} - \frac{i \xi}{p^2 - \xi M_A^2} L_{\mu\nu} \, , \qquad
    \Delta_{a_\mu a_\nu} = \frac{i}{-p^2 + M_A^2} T_{\mu\nu} + \frac{i}{M_A^2} L_{\mu\nu} \, , \nonumber \\
    & \Delta_{a_\mu \chi} = \frac{p_\mu}{M_A (p^2 - \xi M_A^2) } \, , \qquad
    \Delta_{b \chi} = \frac{i M_A}{p^2 - \xi M_A^2} \, , \qquad
    \Delta_{\chi\chi} = \frac{i}{p^2- \xi M_A^2} \, .
    \label{rxi.prop.symm}
\end{align}

The mass eigenstates are expressed in terms of the symmetric basis fields as
\begin{align}
    & b'= b - \frac{1}{\xi} \partial A - M_A \chi \, , \qquad
    A'_\mu = A_\mu - X_\mu \, , 
    \nonumber \\
    & X''_\mu = X_\mu - a_\mu - \frac{1}{M_A} \partial_\mu \chi \, , \qquad
    a'_\mu = a_\mu + \frac{1}{M_A} \partial_\mu \chi \, .
    \label{rxi.vec.fred}
\end{align}

\section{Field UV Dimensions}\label{app:fields.dim}

The functional identities of the theory, once promoted at the quantum level,  
enforce a hierarchy among 1-PI amplitudes.
Some of them in fact are independent
(ancestor amplitudes), others are fixed by the relevant functional identities
and are thus  called descendant amplitudes.

We denote by $\G$ the full vertex functional.
$\G$ can be expanded according to the loop number as
\begin{align}
    \G = \sum_{j=0}^\infty \G^{(j)} \, ,
    \label{loop.vertex.funct}
\end{align}
$\G^{(0)}$ being given by Eq.(\ref{cl.vertex.functional}). 

Since the theory is non-anomalous, the functional identities in Appendix~\ref{app.funct.ids} hold true at the quantum level, provided that a suitable choice of finite counter-terms is carried out order by order in the loop expansion in order to
restore the identities possibly broken by intermediate regularization~\cite{Piguet:1995er}.

We first discuss the relevant functional identities in the symmetric basis.

Eq.(\ref{b.eq}) implies that the $b$-field does not propagate, since
\begin{align}
    \frac{\delta \G^{(j)}}{\delta b } = 0 \, ,\qquad j \geq 1 \, .
    \label{b.j}
\end{align}
The ghost equation (\ref{gh.eq})
becomes at the quantum level
\begin{align}
\frac{\delta \G^{(j)}}{\delta \bar \omega} =  \xi M_A \frac{\delta \G^{(j)}}{\delta \chi^*} \, , \, \qquad j \geq 1
\end{align}
and entails that the dependence on the ghost $\omega$ only happens via the combination
$\chi^{*'} \equiv \chi^* + \xi M_A \omega$.

The constraints ghost and antighost fields are free
by Eqs.(\ref{barc.c.eq}) and (\ref{vect.barc.c.eq}).

The $X$ and $X_\mu$-equations are
\begin{align}
    \frac{\delta \G^{(j)}}{\delta X} = (\square + m^2)
    \frac{\delta \G^{(j)}}{\delta \bar c^*} \, , 
    \qquad
     \frac{\delta \G^{(j)}}{\delta X_\mu} =     \Sigma_{(\xi)}^{\mu\nu} \frac{\delta \G^{(j)}}{\delta \bar c^{\nu *}} \, , 
    \qquad j \geq 1 \,  .
    \label{X.eqs.j}
\end{align}
while the $h$ and $a_\mu$-equations (\ref{h.eq}) and 
(\ref{amu.eq}) become
\begin{align}
    \frac{\delta \G^{(j)}}{\delta h} = 0 \, , \qquad
    \frac{\delta \G^{(j)}}{\delta a_\mu} = 0 \, , \qquad j \geq 1 \, .
    \label{ha.eqs.j}
\end{align}
In order to identify the UV dimension of the fields we need to obtain the relevant functional identities controlling the descendant amplitudes
in the mass eigenstate basis.

By using the chain rule of differentiation we get in the scalar sector 
\begin{align}
    & \frac{\delta \G^{(j)}}{\delta X'}  = \frac{\delta \G^{(j)}}{\delta X} + \frac{\delta \G^{(j)}}{\delta \sigma} = 
    (\square + m^2) \frac{\delta \G^{(j)}}{\delta \bar c^*} + 
    \frac{\delta \G^{(j)}}{\delta \sigma'} \, , 
    \label{X.eq.me}
    \\
    & \frac{\delta \G^{(j)}}{\delta h}  = \frac{\delta \G^{(j)}}{\delta X} + \frac{\delta \G^{(j)}}{\delta \sigma} 
    + \frac{\delta \G^{(j)}}{\delta h }= 
    (\square + m^2) \frac{\delta \G^{(j)}}{\delta \bar c^*} + 
    \frac{\delta \G^{(j)}}{\delta \sigma'} \, ,  \qquad j \geq 1
    \label{h.eq.me}
\end{align}
where use has been made  of the first of Eqs.(\ref{X.eqs.j}) 
and the first of Eqs.(\ref{ha.eqs.j}).
The r.h.s. of Eqs.(\ref{h.eq.me}) is understood to be expressed in terms of the mass eigenstate fields. 

We see from 
Eqs.(\ref{X.eq.me}) and (\ref{h.eq.me}) that insertions of $X'$ and of $h$ are not independent since they are controlled by insertions of the external source
$\bar c^*$ and the field $\sigma'$.

We now consider the vector sector.
\subsection{Landau gauge}

We first consider the $b$-equation.
By applying the chain rule of differentation we get
\begin{align}
    \frac{\delta \G^{(j)}}{\delta b'} & = \frac{\delta \G^{(j)}}{\delta b} + \frac{1}{M_A^2} \partial^\mu \frac{\delta \G^{(j)}}{\delta A_\mu} + \frac{1}{M_A^2} \partial^\mu \frac{\delta \G^{(j)}}{\delta a_\mu} =
    \frac{1}{M_A^2} \partial^\mu \frac{\delta \G^{(j)}}{\delta A''_\mu} \, , \quad j \geq 1 
\end{align}
where we have used Eqs.(\ref{b.j}) and (\ref{ha.eqs.j}).

Again by using the chain rule of differentiation one finds in a similar way
\begin{align}
    & \frac{\delta \G^{(j)}}{\delta X''_\mu} =
    \frac{\delta \G^{(j)}}{\delta X_\mu} + 
    \frac{\delta \G^{(j)}}{\delta A_\mu} = \Sigma_{(0)}^{\mu\nu} \frac{\delta \G^{(j)}}{\delta \bar c^{\nu *}}
    + 
    \frac{\delta \G^{(j)}}{\delta A''_\mu} \, , 
    \label{Xmu.eq.me.0} 
    \\
    &
    \frac{\delta \G^{(j)}}{\delta a'_\mu} =  
    \frac{\delta \G^{(j)}}{\delta A_\mu} + 
    \frac{\delta \G^{(j)}}{\delta X_\mu} +
    \frac{\delta \G^{(j)}}{\delta a_\mu} = 
    \Sigma_{(0)}^{\mu\nu} \frac{\delta \G^{(j)}}{\delta \bar c^{\nu *}}
    + 
    \frac{\delta \G^{(j)}}{\delta A''_\mu} \, , 
    \quad j \geq 1
    \label{amu.eq.me.0} 
\end{align}
by using the second of Eqs.(\ref{X.eqs.j}) 
and the second of Eqs.(\ref{ha.eqs.j}).
We conclude from the above equations
that amplitudes involving 
$X''_\mu$ and $a'_\mu$ are descendant
and are fixed by insertions of 
the external source
$\bar c^*_\mu$ and the quantum field 
$A''_\mu$.

\subsection{$R_\xi$-gauge}

The $b$-equation simply becomes
\begin{align}
\frac{\delta \G^{(j)}}{\delta b'} = 
\frac{\delta \G^{(j)}}{\delta b} = 0 \, , \quad j \geq 1 \, .
\label{b.eq.xi}
\end{align}
The $X_\mu$-equation reads
\begin{align}
    \frac{\delta \G^{(j)}}{\delta X''_\mu} = -\frac{1}{\xi}
    \partial^\mu \frac{\delta \G^{(j)}}{\delta b}  + \frac{\delta \G^{(j)}}{\delta X_\mu} + 
    \frac{\delta \G^{(j)}}{\delta A_\mu} = 
    \Sigma_{(\xi)}^{\mu\nu} \frac{\delta \G^{(j)}}{\delta \bar c^{\nu *}} + 
    \frac{\delta \G^{(j)}}{\delta A_\mu} 
    =   \Sigma_{(\xi)}^{\mu\nu} \frac{\delta \G^{(j)}}{\delta \bar c^{\nu *}} + \frac{\delta \G^{(j)}}{\delta A'_\mu} \, , \quad j \geq 1
    \label{Xmu.eq.me.xi}
\end{align}
where we have used the second of Eqs.(\ref{X.eqs.j}). Notice that
\begin{align}
\frac{\delta \G^{(j)}}{\delta A^\mu} =
\frac{1}{\xi} \partial^\mu 
\frac{\delta \G^{(j)}}{\delta b'} +
\frac{\delta \G^{(j)}}{\delta A'_\mu} = \frac{\delta \G^{(j)}}{\delta A'_\mu}
\end{align}
by Eq.(\ref{b.eq.xi}).

Along the same way one gets the $a'_\mu$-equation
\begin{align}
    \frac{\delta \G^{(j)}}{\delta a'_\mu}  = 
    \Sigma_{(\xi)}^{\mu\nu} \frac{\delta \G^{(j)}}{\delta \bar c^{\nu *}}
    + 
    \frac{\delta \G^{(j)}}{\delta A'_\mu} \, , 
    \quad j \geq 1 \, .
    \label{amu.eq.me.xi}
\end{align}

Again the amplitudes involving $X''_\mu$ and $a'_\mu$ 
are fixed by insertions of 
the external source
$\bar c^*_\mu$ and the quantum field 
$A'_\mu$ by Eqs.(\ref{Xmu.eq.me.xi}) and (\ref{amu.eq.me.xi}).

\subsection{ST identity}

The ST identity Eq.(\ref{sti}) holds true for the full vertex functional $\G$:
\begin{align}
    {\cal S} (\G) =
    \int \d \, \Big [ & \partial_\mu \omega 
    \frac{\delta \G}{\delta A_\mu} + 
    \frac{\delta \G}{\delta \sigma^*}
    \frac{\delta \G}{\delta \sigma} +
    \frac{\delta \G}{\delta \chi^*}
    \frac{\delta \G}{\delta \chi} 
    %\nonumber \\
    %& 
    +
    \frac{\delta \G}{\delta \bar \eta}
    \frac{\delta \G}{\delta \psi} +
    \frac{\delta \G}{\delta \eta}
    \frac{\delta \G}{\delta \bar \psi} 
    + b \frac{\delta \G}{\delta \bar \omega}
     \Big ] = 0 \, .
     \label{sti.full}
\end{align}
Its projection at order $j \geq 1$ in the loop expansion yields
\begin{align}
    {\cal S} (\G)^{(j)} = & {\cal S}_0(\G^{(j)})  \nonumber \\
    &+ 
    \sum_{k=1}^{j-1} 
     \Big [  
    \frac{\delta \G^{(k)}}{\delta \sigma^*}
    \frac{\delta \G^{(j-k)}}{\delta \sigma} +
     \frac{\delta \G^{(k)}}{\delta \chi^*}
    \frac{\delta \G^{(j-k)}}{\delta \chi} +  
    \frac{\delta \G^{(k)}}{\delta \bar \eta}
    \frac{\delta \G^{(j-k)}}{\delta \psi} +
    \frac{\delta \G^{(k)}}{\delta \eta}
    \frac{\delta \G^{(j-k)}}{\delta \bar \psi}    
    \Big ] = 0 \, ,
     \label{sti.full.jth}
\end{align}
where ${\cal S}_0$ is the linearized ST operator
\begin{align}
    {\cal S}_0(\G^{(j)}) = 	
    \int \d  \, \Bigg [ 
    & 
    \partial_\mu \omega 
    \frac{\delta \G^{(j)}}{\delta A_\mu}  + b \frac{\delta \G^{(j)}}{\delta \bar \omega} +
    \frac{\delta \G^{(0)}}{\delta \sigma^*}
    \frac{\delta \G^{(j)}}{\delta \sigma} +
    \frac{\delta \G^{(j)}}{\delta \sigma^*}
    \frac{\delta \G^{(0)}}{\delta \sigma} +
    \frac{\delta \G^{(0)}}{\delta \chi^*}
    \frac{\delta \G^{(j)}}{\delta \chi} +
    \frac{\delta \G^{(j)}}{\delta \chi^*}
    \frac{\delta \G^{(0)}}{\delta \chi}
    \nonumber \\
    & +
    \frac{\delta \G^{(0)}}{\delta \bar \eta}
    \frac{\delta \G^{(j)}}{\delta \psi}
   +
    \frac{\delta \G^{(j)}}{\delta \bar \eta}
    \frac{\delta \G^{(0)}}{\delta \psi} 
    +
    \frac{\delta \G^{(0)}}{\delta \eta}
    \frac{\delta \G^{(j)}}{\delta \bar \psi}    
     +
    \frac{\delta \G^{(j)}}{\delta \eta}
    \frac{\delta \G^{(0)}}{\delta \bar \psi}     
     \Bigg ] = 0 \, .
\end{align}

%
%
%\begin{align}
%    {\cal S}_0(\G^{(j)}) = 	
%    \int \d \, \Big [ 
%    & 
%    \partial_\mu \omega 
%    \frac{\delta \G^{(j)}{\delta A_\mu}  + b \frac{\delta \G}{\delta \bar \omega}
%    %
%    \frac{\delta \G^{(0)}}{\delta \sigma^*}
%    \frac{\delta \G^{(j)}}{\delta \sigma} +
%    %
%    \frac{\delta \G^{(j)}}{\delta \sigma^*}
%    \frac{\delta \G^{(0)}}{\delta \sigma} +
%    %
%    \frac{\delta \G^{(0)}}{\delta \chi^*}
%    \frac{\delta \G^{(j)}}{\delta \chi} +
%    %
%    \frac{\delta \G^{(j)}}{\delta \chi^*}
%    \frac{\delta \G^{(0)}}{\delta \chi}
%    \nonumber \\
%    & +
%    \frac{\delta \G^{(j)}}{\delta \bar \eta}
%    \frac{\delta \G^{(0)}}{\delta \psi}
%   +
%    \frac{\delta \G^{(0)}}{\delta \bar \eta}
%    \frac{\delta \G^{(j)}}{\delta \psi} 
%    +
%    \frac{\delta \G^{(0)}}{\delta \eta}
%    \frac{\delta \G?{(j)}}{\delta \bar \psi}    
%     +
%    \frac{\delta \G^{(j)}}{\delta \eta}
%    \frac{\delta \G?{(0)}}{\delta \bar \psi}     
%     \Big ] = 0 \, .
%\end{align}
%
\subsection{UV dimension of ancestor fields and external sources}
We can now finally obtain the UV dimension of the ancestor fields and
external sources.
By inspection of the propagators and the interaction vertices one sees that
$A'_\mu$ ($A''_\mu$ in Landau gauge),
$\phi$, $\omega$, $\bar \omega$, $c$, $\bar c$, $c_\mu$, $\bar c_\mu$ have UV dimension $1$,
$\psi, \bar \psi$ and 
$\eta, \bar \eta$  have UV dimension $3/2$,
$\sigma^*, \chi^*, \bar c^*$ have UV dimension $2$ and $\bar c^*_\mu$ has UV dimension $1$.

\section{$(\ln,\ln')$-decomposition of the two-point ST identities}\label{app:2pt.sti}

We report here the explicit expression in the Landau gauge of the relevant non-vanishing 2-point amplitudes
evaluated according to the  $\ln,\ln'$-expansion in the number of internal $h$ and $a'_\mu$-lines respectively
in the sectors $(1,0)$ and $(0,1)$.
Evaluation of the Feynman amplitudes is carried out with
the FeynCalc package~\cite{Shtabovenko:2016sxi,Shtabovenko:2020gxv}. We express the amplitudes
on a basis of scalar Passarino-Veltman 2-point functions at $D=4$.

The amplitudes are denoted by the index corresponding to the particular diagram entering into their decomposition.
\begin{align}
 \G^{(1;1,0)}_{A_\mu A_\nu;(i)} & = 
\frac{i \pi^2 M_A^2}{3 v^2 p^4}
\Bigg \{ 
\Big [ (p^2 - M^2)^2 (p^2 g^{\mu\nu} - 4 p^\mu p^\nu) \Big ] 
B_0(p^2,0,M^2)
\nonumber \\
%%%%%%%%
& \qquad \qquad -
\Big [ M^2 p^2 (p^2 + M^2) g^{\mu\nu} - 4 
(M^2 - 2 p^2) p^\mu p^\nu) \Big ] 
B_0(0,M^2,M^2) \Bigg \} \, ,
\nonumber \\
\G^{(1;1,0)}_{A_\mu A_\nu;(ii)} & = 
\frac{i \pi^2 M_A^2}{3 v^2 p^4}
\Bigg \{ 
\Big [ - (p^2 - M^2)^2 p^2 g^{\mu\nu}  + 
(p^4 - 2 M^2 p^2 + 4 M^4) p^\mu p^\nu \Big ] 
B_0(p^2,0,M^2)
\nonumber \\
& \qquad \qquad +
\Big [ M^2 p^2 (p^2 + M^2) g^{\mu\nu} + 2
(p^2-2 M^2) p^\mu p^\nu) \Big ] 
B_0(0,M^2,M^2) \Bigg \} \, , 
\nonumber \\
\G^{(1;1,0)}_{A_\mu A_\nu;(iv)} & =
- \frac{i \pi^2 M^2 M_A^2}{v^2} g^{\mu\nu} B_0(0,M^2,M^2) \ , 
\nonumber \\
%%%%%%
\G^{(1;1,0)}_{A_\nu(-p) \chi(p);(I)} & = 
\frac{\pi^2 M_A p_\nu}{v^2 p^2}
\Bigg \{ 
 - (p^2 - M^2)^2   
B_0(p^2,0,M^2) +
 M^2 (M^2 - p^2) 
B_0(0,M^2,M^2) \Bigg \} \, , 
\nonumber \\
\G^{(1;1,0)}_{A_\nu(-p) \chi(p);(II)} & = 
\frac{\pi^2 M_A M^4 p_\nu}{v^2 p^2}
\Bigg [ 
- B_0(p^2,0,M^2) +
B_0(0,M^2,M^2)
\Bigg ] \, , 
\nonumber \\
\G^{(1;1,1)}_{A_\mu A_\nu;(i)} & = 
\frac{i \pi^2 M_A^2}{3 v^2 p^4}
\Bigg \{ \Big [ 
 M^2 \Big ( p^4 + (M^2 - M_A^2) p^2 \Big ) g^{\mu\nu}
+ 4 M^2 \Big ( 2 p^2 - M^2 + M_A^2 \Big ) p^\mu p^\nu  \Big ] 
B_0(0,M^2,M^2) 
\nonumber \\
& \qquad +
M_A^2 (p^2 - M^2 + M_A^2) (p^2 g^{\mu\nu} - 4 p^\mu p^\nu) B_0(0,M_A^2,M_A^2) 
\nonumber \\
& \qquad - \Big [ \Big ( p^6 -2 (M^2 - 5 M_A^2) p^4 
 + (M^2 -  M_A^2)^2 p^2 \Big ) g^{\mu\nu} 
 \nonumber \\
& \qquad - 4
\Big ( p^4 + (M_A^2 - 2M^2) p^2 + (M^2 - M_A^2)^2 \Big ) p^\mu p^\nu \Big ] B_0(p^2, M^2, M_A^2) \Bigg \} \, ,
\nonumber \\
\G^{(1;1,1)}_{A_\nu(-p) \chi(p) ;(I)} & =
\frac{\pi^2 M_A p_\nu}{v^2 p^2} 
\Bigg \{ 
\Big ( M^2 ( p^2 - M^2 + M_A^2 \Big ) B_0(0,M^2,M^2)
-
\Big ( 
p^2 -M^2 + M_A^2 
\Big ) B_0(0,M_A^2,M_A^2) \nonumber \\
& \qquad + 
\Big ( p^4 -2 (M^2 + M_A^2) p^2 + (M^2 - M_A^2)^2
\Big ) B_0(p^2, M^2, M_A^2 )
\Bigg \} \, .
\label{symm.contr}
\end{align}

The contributions from the $X_\mu$-dependent vertices are
\begin{align}
\G^{(1;1,0)}_{A_\mu A_\nu;(\alpha)} & = -
\frac{i \pi^2 M^2}{v^2} 
\Big [ (4 M_A^2 - 3 M^2) g_{\mu\nu} - 4 (p^2 g_{\mu\nu} - p_\mu p_\nu) \Big ] 
B_0(0,M^2,M^2) \, , \nonumber \\
\G^{(1;1,1)}_{A_\mu A_\nu;(\alpha)}  & =
\frac{i \pi^2 M^2}{v^2} 
\Big [ (4 M_A^2 - 3 M^2) g_{\mu\nu} - 4 (p^2 g_{\mu\nu} - p_\mu p_\nu) \Big ] 
B_0(0,M^2,M^2) \, , \nonumber \\
%%%%%%%%%%%%%%%%
%
\G^{(1;1,0)}_{A_\mu A_\nu;(\beta)} & = 
\frac{8 i \pi^2 M^2 M_A^2}{v^2} g_{\mu\nu}  B_0(0,M^2,M^2) \, , 
\nonumber \\
\G^{(1;1,1)}_{A_\mu A_\nu;(\beta)} & = 
- \frac{8 i \pi^2 M^2 M_A^2}{v^2} g_{\mu\nu}  B_0(0,M^2,M^2)
\, , 
\nonumber \\
\G^{(1;1,0)}_{\chi(p) A_\nu;(\gamma)} & = -
\frac{ \pi^2 M^2 ( 2 M_A^2 - 3 M^2)}{M_A v^2} p_\nu  B_0(0,M^2,M^2) \, , 
\nonumber \\
\G^{(1;1,1)}_{\chi(p) A_\nu;(\gamma)} & = 
\frac{ \pi^2 M^2 ( 2 M_A^2 - 3 M^2)}{M_A v^2} p_\nu  B_0(0,M^2,M^2)  \, , 
\nonumber \\
\G^{(1;1,0)}_{\chi(p) A_\nu;(\delta)} & = 
\frac{4 \pi^2 M^2 M_A}{v^2} p_\nu  B_0(0,M^2,M^2)  \, , 
\nonumber \\
\G^{(1;1,1)}_{\chi(p) A_\nu;(\delta)} & = -
\frac{4 \pi^2 M^2 M_A}{v^2} p_\nu  B_0(0,M^2,M^2) \, .
\label{xmu.contr}
\end{align}

%\bibliography{bibliography_1loop}

\begin{thebibliography}{39}%
\makeatletter
\providecommand \@ifxundefined [1]{%
 \@ifx{#1\undefined}
}%
\providecommand \@ifnum [1]{%
 \ifnum #1\expandafter \@firstoftwo
 \else \expandafter \@secondoftwo
 \fi
}%
\providecommand \@ifx [1]{%
 \ifx #1\expandafter \@firstoftwo
 \else \expandafter \@secondoftwo
 \fi
}%
\providecommand \natexlab [1]{#1}%
\providecommand \enquote  [1]{``#1''}%
\providecommand \bibnamefont  [1]{#1}%
\providecommand \bibfnamefont [1]{#1}%
\providecommand \citenamefont [1]{#1}%
\providecommand \href@noop [0]{\@secondoftwo}%
\providecommand \href [0]{\begingroup \@sanitize@url \@href}%
\providecommand \@href[1]{\@@startlink{#1}\@@href}%
\providecommand \@@href[1]{\endgroup#1\@@endlink}%
\providecommand \@sanitize@url [0]{\catcode `\\12\catcode `\$12\catcode
  `\&12\catcode `\#12\catcode `\^12\catcode `\_12\catcode `\%12\relax}%
\providecommand \@@startlink[1]{}%
\providecommand \@@endlink[0]{}%
\providecommand \url  [0]{\begingroup\@sanitize@url \@url }%
\providecommand \@url [1]{\endgroup\@href {#1}{\urlprefix }}%
\providecommand \urlprefix  [0]{URL }%
\providecommand \Eprint [0]{\href }%
\providecommand \doibase [0]{http://dx.doi.org/}%
\providecommand \selectlanguage [0]{\@gobble}%
\providecommand \bibinfo  [0]{\@secondoftwo}%
\providecommand \bibfield  [0]{\@secondoftwo}%
\providecommand \translation [1]{[#1]}%
\providecommand \BibitemOpen [0]{}%
\providecommand \bibitemStop [0]{}%
\providecommand \bibitemNoStop [0]{.\EOS\space}%
\providecommand \EOS [0]{\spacefactor3000\relax}%
\providecommand \BibitemShut  [1]{\csname bibitem#1\endcsname}%
\let\auto@bib@innerbib\@empty
%</preamble>
\bibitem [{\citenamefont {B\'elusca-Maito}\ \emph {et~al.}(2023)\citenamefont
  {B\'elusca-Maito}, \citenamefont {Ilakovac}, \citenamefont {Kuehler},
  \citenamefont {Mador-Bovzinovic}, \citenamefont {Stoeckinger},\ and\
  \citenamefont {Weisswange}}]{Belusca-Maito:2023wah}%
  \BibitemOpen
  \bibfield  {author} {\bibinfo {author} {\bibfnamefont {H.}~\bibnamefont
  {B\'elusca-Maito}}, \bibinfo {author} {\bibfnamefont {A.}~\bibnamefont
  {Ilakovac}}, \bibinfo {author} {\bibfnamefont {P.}~\bibnamefont {Kuehler}},
  \bibinfo {author} {\bibfnamefont {M.}~\bibnamefont {Mador-Bovzinovic}},
  \bibinfo {author} {\bibfnamefont {D.}~\bibnamefont {Stoeckinger}}, \ and\
  \bibinfo {author} {\bibfnamefont {M.}~\bibnamefont {Weisswange}},\ }\href
  {\doibase 10.3390/sym15030622} {\bibfield  {journal} {\bibinfo  {journal}
  {Symmetry}\ }\textbf {\bibinfo {volume} {15}},\ \bibinfo {pages} {622}
  (\bibinfo {year} {2023})},\ \Eprint {http://arxiv.org/abs/2303.09120}
  {arXiv:2303.09120 [hep-ph]} \BibitemShut {NoStop}%
\bibitem [{\citenamefont {Ferrari}\ \emph {et~al.}(2000)\citenamefont
  {Ferrari}, \citenamefont {Grassi},\ and\ \citenamefont
  {Quadri}}]{Ferrari:1999nj}%
  \BibitemOpen
  \bibfield  {author} {\bibinfo {author} {\bibfnamefont {R.}~\bibnamefont
  {Ferrari}}, \bibinfo {author} {\bibfnamefont {P.~A.}\ \bibnamefont {Grassi}},
  \ and\ \bibinfo {author} {\bibfnamefont {A.}~\bibnamefont {Quadri}},\ }\href
  {\doibase 10.1016/S0370-2693(99)01452-5} {\bibfield  {journal} {\bibinfo
  {journal} {Phys. Lett.}\ }\textbf {\bibinfo {volume} {B472}},\ \bibinfo
  {pages} {346} (\bibinfo {year} {2000})},\ \Eprint
  {http://arxiv.org/abs/hep-th/9905192} {arXiv:hep-th/9905192 [hep-th]}
  \BibitemShut {NoStop}%
%%CITATION = HEP-TH/9905192;%%
\bibitem [{\citenamefont {Ferrari}\ and\ \citenamefont
  {Grassi}(1999)}]{Ferrari:1998jy}%
  \BibitemOpen
  \bibfield  {author} {\bibinfo {author} {\bibfnamefont {R.}~\bibnamefont
  {Ferrari}}\ and\ \bibinfo {author} {\bibfnamefont {P.~A.}\ \bibnamefont
  {Grassi}},\ }\href {\doibase 10.1103/PhysRevD.60.065010} {\bibfield
  {journal} {\bibinfo  {journal} {Phys. Rev.}\ }\textbf {\bibinfo {volume}
  {D60}},\ \bibinfo {pages} {065010} (\bibinfo {year} {1999})},\ \Eprint
  {http://arxiv.org/abs/hep-th/9807191} {arXiv:hep-th/9807191 [hep-th]}
  \BibitemShut {NoStop}%
%%CITATION = HEP-TH/9807191;%%
\bibitem [{\citenamefont {Grassi}\ \emph {et~al.}(2001)\citenamefont {Grassi},
  \citenamefont {Hurth},\ and\ \citenamefont {Steinhauser}}]{Grassi:1999tp}%
  \BibitemOpen
  \bibfield  {author} {\bibinfo {author} {\bibfnamefont {P.~A.}\ \bibnamefont
  {Grassi}}, \bibinfo {author} {\bibfnamefont {T.}~\bibnamefont {Hurth}}, \
  and\ \bibinfo {author} {\bibfnamefont {M.}~\bibnamefont {Steinhauser}},\
  }\href {\doibase 10.1006/aphy.2001.6117} {\bibfield  {journal} {\bibinfo
  {journal} {Annals Phys.}\ }\textbf {\bibinfo {volume} {288}},\ \bibinfo
  {pages} {197} (\bibinfo {year} {2001})},\ \Eprint
  {http://arxiv.org/abs/hep-ph/9907426} {arXiv:hep-ph/9907426 [hep-ph]}
  \BibitemShut {NoStop}%
%%CITATION = HEP-PH/9907426;%%
\bibitem [{\citenamefont {Hollik}\ \emph {et~al.}(1999)\citenamefont {Hollik},
  \citenamefont {Kraus},\ and\ \citenamefont {Stockinger}}]{Hollik:1999xh}%
  \BibitemOpen
  \bibfield  {author} {\bibinfo {author} {\bibfnamefont {W.}~\bibnamefont
  {Hollik}}, \bibinfo {author} {\bibfnamefont {E.}~\bibnamefont {Kraus}}, \
  and\ \bibinfo {author} {\bibfnamefont {D.}~\bibnamefont {Stockinger}},\
  }\href {\doibase 10.1007/s100520050642} {\bibfield  {journal} {\bibinfo
  {journal} {Eur. Phys. J. C}\ }\textbf {\bibinfo {volume} {11}},\ \bibinfo
  {pages} {365} (\bibinfo {year} {1999})},\ \Eprint
  {http://arxiv.org/abs/hep-ph/9907393} {arXiv:hep-ph/9907393} \BibitemShut
  {NoStop}%
\bibitem [{\citenamefont {Hollik}\ and\ \citenamefont
  {Stockinger}(2001)}]{Hollik:2001cz}%
  \BibitemOpen
  \bibfield  {author} {\bibinfo {author} {\bibfnamefont {W.}~\bibnamefont
  {Hollik}}\ and\ \bibinfo {author} {\bibfnamefont {D.}~\bibnamefont
  {Stockinger}},\ }\href {\doibase 10.1007/s100520100651} {\bibfield  {journal}
  {\bibinfo  {journal} {Eur. Phys. J. C}\ }\textbf {\bibinfo {volume} {20}},\
  \bibinfo {pages} {105} (\bibinfo {year} {2001})},\ \Eprint
  {http://arxiv.org/abs/hep-ph/0103009} {arXiv:hep-ph/0103009} \BibitemShut
  {NoStop}%
\bibitem [{\citenamefont {Stockinger}(2005)}]{Stockinger:2005gx}%
  \BibitemOpen
  \bibfield  {author} {\bibinfo {author} {\bibfnamefont {D.}~\bibnamefont
  {Stockinger}},\ }\href {\doibase 10.1088/1126-6708/2005/03/076} {\bibfield
  {journal} {\bibinfo  {journal} {JHEP}\ }\textbf {\bibinfo {volume} {03}},\
  \bibinfo {pages} {076} (\bibinfo {year} {2005})},\ \Eprint
  {http://arxiv.org/abs/hep-ph/0503129} {arXiv:hep-ph/0503129} \BibitemShut
  {NoStop}%
\bibitem [{\citenamefont {Fischer}\ \emph {et~al.}(2004)\citenamefont
  {Fischer}, \citenamefont {Hollik}, \citenamefont {Roth},\ and\ \citenamefont
  {Stockinger}}]{Fischer:2003cb}%
  \BibitemOpen
  \bibfield  {author} {\bibinfo {author} {\bibfnamefont {I.}~\bibnamefont
  {Fischer}}, \bibinfo {author} {\bibfnamefont {W.}~\bibnamefont {Hollik}},
  \bibinfo {author} {\bibfnamefont {M.}~\bibnamefont {Roth}}, \ and\ \bibinfo
  {author} {\bibfnamefont {D.}~\bibnamefont {Stockinger}},\ }\href {\doibase
  10.1103/PhysRevD.69.015004} {\bibfield  {journal} {\bibinfo  {journal} {Phys.
  Rev. D}\ }\textbf {\bibinfo {volume} {69}},\ \bibinfo {pages} {015004}
  (\bibinfo {year} {2004})},\ \Eprint {http://arxiv.org/abs/hep-ph/0310191}
  {arXiv:hep-ph/0310191} \BibitemShut {NoStop}%
\bibitem [{\citenamefont {B\'elusca-Maito}\ \emph {et~al.}(2020)\citenamefont
  {B\'elusca-Maito}, \citenamefont {Ilakovac}, \citenamefont
  {Mador-Bovzinovic},\ and\ \citenamefont
  {St\"ockinger}}]{Belusca-Maito:2020ala}%
  \BibitemOpen
  \bibfield  {author} {\bibinfo {author} {\bibfnamefont {H.}~\bibnamefont
  {B\'elusca-Maito}}, \bibinfo {author} {\bibfnamefont {A.}~\bibnamefont
  {Ilakovac}}, \bibinfo {author} {\bibfnamefont {M.}~\bibnamefont
  {Mador-Bovzinovic}}, \ and\ \bibinfo {author} {\bibfnamefont
  {D.}~\bibnamefont {St\"ockinger}},\ }\href {\doibase 10.1007/JHEP08(2020)024}
  {\bibfield  {journal} {\bibinfo  {journal} {JHEP}\ }\textbf {\bibinfo
  {volume} {08}},\ \bibinfo {pages} {024} (\bibinfo {year} {2020})},\ \Eprint
  {http://arxiv.org/abs/2004.14398} {arXiv:2004.14398 [hep-ph]} \BibitemShut
  {NoStop}%
\bibitem [{\citenamefont {Cornella}\ \emph {et~al.}(2023)\citenamefont
  {Cornella}, \citenamefont {Feruglio},\ and\ \citenamefont
  {Vecchi}}]{Cornella:2022hkc}%
  \BibitemOpen
  \bibfield  {author} {\bibinfo {author} {\bibfnamefont {C.}~\bibnamefont
  {Cornella}}, \bibinfo {author} {\bibfnamefont {F.}~\bibnamefont {Feruglio}},
  \ and\ \bibinfo {author} {\bibfnamefont {L.}~\bibnamefont {Vecchi}},\ }\href
  {\doibase 10.1007/JHEP02(2023)244} {\bibfield  {journal} {\bibinfo  {journal}
  {JHEP}\ }\textbf {\bibinfo {volume} {02}},\ \bibinfo {pages} {244} (\bibinfo
  {year} {2023})},\ \Eprint {http://arxiv.org/abs/2205.10381} {arXiv:2205.10381
  [hep-ph]} \BibitemShut {NoStop}%
\bibitem [{\citenamefont {Frohlich}\ \emph {et~al.}(1980)\citenamefont
  {Frohlich}, \citenamefont {Morchio},\ and\ \citenamefont
  {Strocchi}}]{Frohlich:1980gj}%
  \BibitemOpen
  \bibfield  {author} {\bibinfo {author} {\bibfnamefont {J.}~\bibnamefont
  {Frohlich}}, \bibinfo {author} {\bibfnamefont {G.}~\bibnamefont {Morchio}}, \
  and\ \bibinfo {author} {\bibfnamefont {F.}~\bibnamefont {Strocchi}},\ }\href
  {\doibase 10.1016/0370-2693(80)90594-8} {\bibfield  {journal} {\bibinfo
  {journal} {Phys. Lett. B}\ }\textbf {\bibinfo {volume} {97}},\ \bibinfo
  {pages} {249} (\bibinfo {year} {1980})}\BibitemShut {NoStop}%
\bibitem [{\citenamefont {Frohlich}\ \emph {et~al.}(1981)\citenamefont
  {Frohlich}, \citenamefont {Morchio},\ and\ \citenamefont
  {Strocchi}}]{Frohlich:1981yi}%
  \BibitemOpen
  \bibfield  {author} {\bibinfo {author} {\bibfnamefont {J.}~\bibnamefont
  {Frohlich}}, \bibinfo {author} {\bibfnamefont {G.}~\bibnamefont {Morchio}}, \
  and\ \bibinfo {author} {\bibfnamefont {F.}~\bibnamefont {Strocchi}},\ }\href
  {\doibase 10.1016/0550-3213(81)90448-X} {\bibfield  {journal} {\bibinfo
  {journal} {Nucl. Phys. B}\ }\textbf {\bibinfo {volume} {190}},\ \bibinfo
  {pages} {553} (\bibinfo {year} {1981})}\BibitemShut {NoStop}%
\bibitem [{\citenamefont {Clark}(1975)}]{Clark:1974eq}%
  \BibitemOpen
  \bibfield  {author} {\bibinfo {author} {\bibfnamefont {T.~E.}\ \bibnamefont
  {Clark}},\ }\href {\doibase 10.1016/0550-3213(75)90658-6} {\bibfield
  {journal} {\bibinfo  {journal} {Nucl. Phys. B}\ }\textbf {\bibinfo {volume}
  {90}},\ \bibinfo {pages} {484} (\bibinfo {year} {1975})}\BibitemShut
  {NoStop}%
\bibitem [{\citenamefont {Maas}\ \emph {et~al.}(2019)\citenamefont {Maas},
  \citenamefont {Sondenheimer},\ and\ \citenamefont {T\"orek}}]{Maas:2017xzh}%
  \BibitemOpen
  \bibfield  {author} {\bibinfo {author} {\bibfnamefont {A.}~\bibnamefont
  {Maas}}, \bibinfo {author} {\bibfnamefont {R.}~\bibnamefont {Sondenheimer}},
  \ and\ \bibinfo {author} {\bibfnamefont {P.}~\bibnamefont {T\"orek}},\ }\href
  {\doibase 10.1016/j.aop.2019.01.010} {\bibfield  {journal} {\bibinfo
  {journal} {Annals Phys.}\ }\textbf {\bibinfo {volume} {402}},\ \bibinfo
  {pages} {18} (\bibinfo {year} {2019})},\ \Eprint
  {http://arxiv.org/abs/1709.07477} {arXiv:1709.07477 [hep-ph]} \BibitemShut
  {NoStop}%
\bibitem [{\citenamefont {Maas}(2019)}]{Maas:2017wzi}%
  \BibitemOpen
  \bibfield  {author} {\bibinfo {author} {\bibfnamefont {A.}~\bibnamefont
  {Maas}},\ }\href {\doibase 10.1016/j.ppnp.2019.02.003} {\bibfield  {journal}
  {\bibinfo  {journal} {Prog. Part. Nucl. Phys.}\ }\textbf {\bibinfo {volume}
  {106}},\ \bibinfo {pages} {132} (\bibinfo {year} {2019})},\ \Eprint
  {http://arxiv.org/abs/1712.04721} {arXiv:1712.04721 [hep-ph]} \BibitemShut
  {NoStop}%
\bibitem [{\citenamefont {Dudal}\ \emph {et~al.}(2023)\citenamefont {Dudal},
  \citenamefont {van Egmond}, \citenamefont {Peruzzo},\ and\ \citenamefont
  {Sorella}}]{Dudal:2023jsu}%
  \BibitemOpen
  \bibfield  {author} {\bibinfo {author} {\bibfnamefont {D.}~\bibnamefont
  {Dudal}}, \bibinfo {author} {\bibfnamefont {D.~M.}\ \bibnamefont {van
  Egmond}}, \bibinfo {author} {\bibfnamefont {G.}~\bibnamefont {Peruzzo}}, \
  and\ \bibinfo {author} {\bibfnamefont {S.~P.}\ \bibnamefont {Sorella}},\
  }\href {\doibase 10.1140/epjc/s10052-023-12272-6} {\bibfield  {journal}
  {\bibinfo  {journal} {Eur. Phys. J. C}\ }\textbf {\bibinfo {volume} {83}},\
  \bibinfo {pages} {1091} (\bibinfo {year} {2023})},\ \Eprint
  {http://arxiv.org/abs/2309.16776} {arXiv:2309.16776 [hep-th]} \BibitemShut
  {NoStop}%
\bibitem [{\citenamefont {Dudal}\ \emph {et~al.}(2022)\citenamefont {Dudal},
  \citenamefont {van Egmond}, \citenamefont {Justo}, \citenamefont {Peruzzo},\
  and\ \citenamefont {Sorella}}]{Dudal:2021dec}%
  \BibitemOpen
  \bibfield  {author} {\bibinfo {author} {\bibfnamefont {D.}~\bibnamefont
  {Dudal}}, \bibinfo {author} {\bibfnamefont {D.~M.}\ \bibnamefont {van
  Egmond}}, \bibinfo {author} {\bibfnamefont {I.~F.}\ \bibnamefont {Justo}},
  \bibinfo {author} {\bibfnamefont {G.}~\bibnamefont {Peruzzo}}, \ and\
  \bibinfo {author} {\bibfnamefont {S.~P.}\ \bibnamefont {Sorella}},\ }\href
  {\doibase 10.1103/PhysRevD.105.065018} {\bibfield  {journal} {\bibinfo
  {journal} {Phys. Rev. D}\ }\textbf {\bibinfo {volume} {105}},\ \bibinfo
  {pages} {065018} (\bibinfo {year} {2022})},\ \Eprint
  {http://arxiv.org/abs/2111.11958} {arXiv:2111.11958 [hep-th]} \BibitemShut
  {NoStop}%
\bibitem [{\citenamefont {Dudal}\ \emph
  {et~al.}(2021{\natexlab{a}})\citenamefont {Dudal}, \citenamefont {Peruzzo},\
  and\ \citenamefont {Sorella}}]{Dudal:2021pvw}%
  \BibitemOpen
  \bibfield  {author} {\bibinfo {author} {\bibfnamefont {D.}~\bibnamefont
  {Dudal}}, \bibinfo {author} {\bibfnamefont {G.}~\bibnamefont {Peruzzo}}, \
  and\ \bibinfo {author} {\bibfnamefont {S.~P.}\ \bibnamefont {Sorella}},\
  }\href {\doibase 10.1007/JHEP10(2021)039} {\bibfield  {journal} {\bibinfo
  {journal} {JHEP}\ }\textbf {\bibinfo {volume} {10}},\ \bibinfo {pages} {039}
  (\bibinfo {year} {2021}{\natexlab{a}})},\ \Eprint
  {http://arxiv.org/abs/2105.11011} {arXiv:2105.11011 [hep-th]} \BibitemShut
  {NoStop}%
\bibitem [{\citenamefont {Dudal}\ \emph
  {et~al.}(2021{\natexlab{b}})\citenamefont {Dudal}, \citenamefont {van
  Egmond}, \citenamefont {Guimaraes}, \citenamefont {Palhares}, \citenamefont
  {Peruzzo},\ and\ \citenamefont {Sorella}}]{Dudal:2020uwb}%
  \BibitemOpen
  \bibfield  {author} {\bibinfo {author} {\bibfnamefont {D.}~\bibnamefont
  {Dudal}}, \bibinfo {author} {\bibfnamefont {D.~M.}\ \bibnamefont {van
  Egmond}}, \bibinfo {author} {\bibfnamefont {M.~S.}\ \bibnamefont
  {Guimaraes}}, \bibinfo {author} {\bibfnamefont {L.~F.}\ \bibnamefont
  {Palhares}}, \bibinfo {author} {\bibfnamefont {G.}~\bibnamefont {Peruzzo}}, \
  and\ \bibinfo {author} {\bibfnamefont {S.~P.}\ \bibnamefont {Sorella}},\
  }\href {\doibase 10.1140/epjc/s10052-021-09008-9} {\bibfield  {journal}
  {\bibinfo  {journal} {Eur. Phys. J. C}\ }\textbf {\bibinfo {volume} {81}},\
  \bibinfo {pages} {222} (\bibinfo {year} {2021}{\natexlab{b}})},\ \Eprint
  {http://arxiv.org/abs/2008.07813} {arXiv:2008.07813 [hep-th]} \BibitemShut
  {NoStop}%
\bibitem [{\citenamefont {Dudal}\ \emph {et~al.}(2020)\citenamefont {Dudal},
  \citenamefont {van Egmond}, \citenamefont {Guimaraes}, \citenamefont
  {Holanda}, \citenamefont {Palhares}, \citenamefont {Peruzzo},\ and\
  \citenamefont {Sorella}}]{Dudal:2019pyg}%
  \BibitemOpen
  \bibfield  {author} {\bibinfo {author} {\bibfnamefont {D.}~\bibnamefont
  {Dudal}}, \bibinfo {author} {\bibfnamefont {D.~M.}\ \bibnamefont {van
  Egmond}}, \bibinfo {author} {\bibfnamefont {M.~S.}\ \bibnamefont
  {Guimaraes}}, \bibinfo {author} {\bibfnamefont {O.}~\bibnamefont {Holanda}},
  \bibinfo {author} {\bibfnamefont {L.~F.}\ \bibnamefont {Palhares}}, \bibinfo
  {author} {\bibfnamefont {G.}~\bibnamefont {Peruzzo}}, \ and\ \bibinfo
  {author} {\bibfnamefont {S.~P.}\ \bibnamefont {Sorella}},\ }\href {\doibase
  10.1007/JHEP02(2020)188} {\bibfield  {journal} {\bibinfo  {journal} {JHEP}\
  }\textbf {\bibinfo {volume} {02}},\ \bibinfo {pages} {188} (\bibinfo {year}
  {2020})},\ \Eprint {http://arxiv.org/abs/1912.11390} {arXiv:1912.11390
  [hep-th]} \BibitemShut {NoStop}%
\bibitem [{\citenamefont {Kugo}\ and\ \citenamefont
  {Ojima}(1978)}]{Kugo:1977zq}%
  \BibitemOpen
  \bibfield  {author} {\bibinfo {author} {\bibfnamefont {T.}~\bibnamefont
  {Kugo}}\ and\ \bibinfo {author} {\bibfnamefont {I.}~\bibnamefont {Ojima}},\
  }\href {\doibase 10.1016/0370-2693(78)90765-7} {\bibfield  {journal}
  {\bibinfo  {journal} {Phys. Lett.}\ }\textbf {\bibinfo {volume} {B73}},\
  \bibinfo {pages} {459} (\bibinfo {year} {1978})}\BibitemShut {NoStop}%
%%CITATION = PHLTA,B73,459;%%
\bibitem [{\citenamefont {Curci}\ and\ \citenamefont
  {Ferrari}(1976)}]{Curci:1976yb}%
  \BibitemOpen
  \bibfield  {author} {\bibinfo {author} {\bibfnamefont {G.}~\bibnamefont
  {Curci}}\ and\ \bibinfo {author} {\bibfnamefont {R.}~\bibnamefont
  {Ferrari}},\ }\href {\doibase 10.1007/BF02730284} {\bibfield  {journal}
  {\bibinfo  {journal} {Nuovo Cim.}\ }\textbf {\bibinfo {volume} {A35}},\
  \bibinfo {pages} {273} (\bibinfo {year} {1976})}\BibitemShut {NoStop}%
%%CITATION = NUCIA,A35,273;%%
\bibitem [{\citenamefont {Becchi}\ \emph {et~al.}(1974)\citenamefont {Becchi},
  \citenamefont {Rouet},\ and\ \citenamefont {Stora}}]{Becchi:1974xu}%
  \BibitemOpen
  \bibfield  {author} {\bibinfo {author} {\bibfnamefont {C.}~\bibnamefont
  {Becchi}}, \bibinfo {author} {\bibfnamefont {A.}~\bibnamefont {Rouet}}, \
  and\ \bibinfo {author} {\bibfnamefont {R.}~\bibnamefont {Stora}},\ }\href
  {\doibase 10.1016/0370-2693(74)90058-6} {\bibfield  {journal} {\bibinfo
  {journal} {Phys. Lett.}\ }\textbf {\bibinfo {volume} {B52}},\ \bibinfo
  {pages} {344} (\bibinfo {year} {1974})}\BibitemShut {NoStop}%
%%CITATION = PHLTA,B52,344;%%
\bibitem [{\citenamefont {Binosi}\ and\ \citenamefont
  {Quadri}(2022)}]{Binosi:2022ycu}%
  \BibitemOpen
  \bibfield  {author} {\bibinfo {author} {\bibfnamefont {D.}~\bibnamefont
  {Binosi}}\ and\ \bibinfo {author} {\bibfnamefont {A.}~\bibnamefont
  {Quadri}},\ }\href {\doibase 10.1103/PhysRevD.106.065022} {\bibfield
  {journal} {\bibinfo  {journal} {Phys. Rev. D}\ }\textbf {\bibinfo {volume}
  {106}},\ \bibinfo {pages} {065022} (\bibinfo {year} {2022})},\ \Eprint
  {http://arxiv.org/abs/2206.00894} {arXiv:2206.00894 [hep-th]} \BibitemShut
  {NoStop}%
\bibitem [{\citenamefont {Piguet}\ and\ \citenamefont
  {Sorella}(1995)}]{Piguet:1995er}%
  \BibitemOpen
  \bibfield  {author} {\bibinfo {author} {\bibfnamefont {O.}~\bibnamefont
  {Piguet}}\ and\ \bibinfo {author} {\bibfnamefont {S.~P.}\ \bibnamefont
  {Sorella}},\ }\href {\doibase 10.1007/978-3-540-49192-7} {\bibfield
  {journal} {\bibinfo  {journal} {Lect. Notes Phys. Monogr.}\ }\textbf
  {\bibinfo {volume} {28}},\ \bibinfo {pages} {1} (\bibinfo {year}
  {1995})}\BibitemShut {NoStop}%
%%CITATION = INSPIRE-405127;%%
\bibitem [{\citenamefont {Quadri}(2017)}]{Quadri:2016wwl}%
  \BibitemOpen
  \bibfield  {author} {\bibinfo {author} {\bibfnamefont {A.}~\bibnamefont
  {Quadri}},\ }\href {\doibase 10.1142/S0217751X17500890} {\bibfield  {journal}
  {\bibinfo  {journal} {Int. J. Mod. Phys.}\ }\textbf {\bibinfo {volume}
  {A32}},\ \bibinfo {pages} {1750089} (\bibinfo {year} {2017})},\ \Eprint
  {http://arxiv.org/abs/1610.00150} {arXiv:1610.00150 [hep-th]} \BibitemShut
  {NoStop}%
%%CITATION = ARXIV:1610.00150;%%
\bibitem [{\citenamefont {Binosi}\ and\ \citenamefont
  {Quadri}(2018)}]{Binosi:2017ubk}%
  \BibitemOpen
  \bibfield  {author} {\bibinfo {author} {\bibfnamefont {D.}~\bibnamefont
  {Binosi}}\ and\ \bibinfo {author} {\bibfnamefont {A.}~\bibnamefont
  {Quadri}},\ }\href {\doibase 10.1007/JHEP04(2018)050} {\bibfield  {journal}
  {\bibinfo  {journal} {JHEP}\ }\textbf {\bibinfo {volume} {04}},\ \bibinfo
  {pages} {050} (\bibinfo {year} {2018})},\ \Eprint
  {http://arxiv.org/abs/1709.09937} {arXiv:1709.09937 [hep-th]} \BibitemShut
  {NoStop}%
\bibitem [{\citenamefont {Binosi}\ and\ \citenamefont
  {Quadri}(2019)}]{Binosi:2019olm}%
  \BibitemOpen
  \bibfield  {author} {\bibinfo {author} {\bibfnamefont {D.}~\bibnamefont
  {Binosi}}\ and\ \bibinfo {author} {\bibfnamefont {A.}~\bibnamefont
  {Quadri}},\ }\href {\doibase 10.1007/JHEP09(2019)032} {\bibfield  {journal}
  {\bibinfo  {journal} {JHEP}\ }\textbf {\bibinfo {volume} {09}},\ \bibinfo
  {pages} {032} (\bibinfo {year} {2019})},\ \Eprint
  {http://arxiv.org/abs/1904.06692} {arXiv:1904.06692 [hep-ph]} \BibitemShut
  {NoStop}%
%%CITATION = ARXIV:1904.06692;%%
\bibitem [{\citenamefont {Binosi}\ and\ \citenamefont
  {Quadri}(2020{\natexlab{a}})}]{Binosi:2019nwz}%
  \BibitemOpen
  \bibfield  {author} {\bibinfo {author} {\bibfnamefont {D.}~\bibnamefont
  {Binosi}}\ and\ \bibinfo {author} {\bibfnamefont {A.}~\bibnamefont
  {Quadri}},\ }\href {\doibase 10.1140/epjc/s10052-020-8349-0} {\bibfield
  {journal} {\bibinfo  {journal} {Eur. Phys. J. C}\ }\textbf {\bibinfo {volume}
  {80}},\ \bibinfo {pages} {807} (\bibinfo {year} {2020}{\natexlab{a}})},\
  \Eprint {http://arxiv.org/abs/1904.06693} {arXiv:1904.06693 [hep-ph]}
  \BibitemShut {NoStop}%
\bibitem [{\citenamefont {Binosi}\ and\ \citenamefont
  {Quadri}(2020{\natexlab{b}})}]{Binosi:2020unh}%
  \BibitemOpen
  \bibfield  {author} {\bibinfo {author} {\bibfnamefont {D.}~\bibnamefont
  {Binosi}}\ and\ \bibinfo {author} {\bibfnamefont {A.}~\bibnamefont
  {Quadri}},\ }\href {\doibase 10.1007/JHEP05(2020)141} {\bibfield  {journal}
  {\bibinfo  {journal} {JHEP}\ }\textbf {\bibinfo {volume} {05}},\ \bibinfo
  {pages} {141} (\bibinfo {year} {2020}{\natexlab{b}})},\ \Eprint
  {http://arxiv.org/abs/2001.07430} {arXiv:2001.07430 [hep-ph]} \BibitemShut
  {NoStop}%
\bibitem [{\citenamefont {Barnich}\ \emph
  {et~al.}(2000{\natexlab{a}})\citenamefont {Barnich}, \citenamefont {Brandt},\
  and\ \citenamefont {Henneaux}}]{Barnich:2000zw}%
  \BibitemOpen
  \bibfield  {author} {\bibinfo {author} {\bibfnamefont {G.}~\bibnamefont
  {Barnich}}, \bibinfo {author} {\bibfnamefont {F.}~\bibnamefont {Brandt}}, \
  and\ \bibinfo {author} {\bibfnamefont {M.}~\bibnamefont {Henneaux}},\ }\href
  {\doibase 10.1016/S0370-1573(00)00049-1} {\bibfield  {journal} {\bibinfo
  {journal} {Phys. Rept.}\ }\textbf {\bibinfo {volume} {338}},\ \bibinfo
  {pages} {439} (\bibinfo {year} {2000}{\natexlab{a}})},\ \Eprint
  {http://arxiv.org/abs/hep-th/0002245} {arXiv:hep-th/0002245 [hep-th]}
  \BibitemShut {NoStop}%
%%CITATION = HEP-TH/0002245;%%
\bibitem [{\citenamefont {Barnich}\ \emph
  {et~al.}(2000{\natexlab{b}})\citenamefont {Barnich}, \citenamefont
  {Henneaux}, \citenamefont {Hurth},\ and\ \citenamefont
  {Skenderis}}]{Barnich:1999cy}%
  \BibitemOpen
  \bibfield  {author} {\bibinfo {author} {\bibfnamefont {G.}~\bibnamefont
  {Barnich}}, \bibinfo {author} {\bibfnamefont {M.}~\bibnamefont {Henneaux}},
  \bibinfo {author} {\bibfnamefont {T.}~\bibnamefont {Hurth}}, \ and\ \bibinfo
  {author} {\bibfnamefont {K.}~\bibnamefont {Skenderis}},\ }\href {\doibase
  10.1016/S0370-2693(00)01087-X} {\bibfield  {journal} {\bibinfo  {journal}
  {Phys. Lett. B}\ }\textbf {\bibinfo {volume} {492}},\ \bibinfo {pages} {376}
  (\bibinfo {year} {2000}{\natexlab{b}})},\ \Eprint
  {http://arxiv.org/abs/hep-th/9910201} {arXiv:hep-th/9910201} \BibitemShut
  {NoStop}%
\bibitem [{\citenamefont {Gomis}\ \emph {et~al.}(1995)\citenamefont {Gomis},
  \citenamefont {Paris},\ and\ \citenamefont {Samuel}}]{Gomis:1994he}%
  \BibitemOpen
  \bibfield  {author} {\bibinfo {author} {\bibfnamefont {J.}~\bibnamefont
  {Gomis}}, \bibinfo {author} {\bibfnamefont {J.}~\bibnamefont {Paris}}, \ and\
  \bibinfo {author} {\bibfnamefont {S.}~\bibnamefont {Samuel}},\ }\href
  {\doibase 10.1016/0370-1573(94)00112-G} {\bibfield  {journal} {\bibinfo
  {journal} {Phys. Rept.}\ }\textbf {\bibinfo {volume} {259}},\ \bibinfo
  {pages} {1} (\bibinfo {year} {1995})},\ \Eprint
  {http://arxiv.org/abs/hep-th/9412228} {arXiv:hep-th/9412228 [hep-th]}
  \BibitemShut {NoStop}%
%%CITATION = HEP-TH/9412228;%%
\bibitem [{\citenamefont {Quadri}(2002)}]{Quadri:2002nh}%
  \BibitemOpen
  \bibfield  {author} {\bibinfo {author} {\bibfnamefont {A.}~\bibnamefont
  {Quadri}},\ }\href {\doibase 10.1088/1126-6708/2002/05/051} {\bibfield
  {journal} {\bibinfo  {journal} {JHEP}\ }\textbf {\bibinfo {volume} {05}},\
  \bibinfo {pages} {051} (\bibinfo {year} {2002})},\ \Eprint
  {http://arxiv.org/abs/hep-th/0201122} {arXiv:hep-th/0201122 [hep-th]}
  \BibitemShut {NoStop}%
%%CITATION = HEP-TH/0201122;%%
\bibitem [{\citenamefont {Anselmi}(2018)}]{Anselmi:2018kgz}%
  \BibitemOpen
  \bibfield  {author} {\bibinfo {author} {\bibfnamefont {D.}~\bibnamefont
  {Anselmi}},\ }\href {\doibase 10.1007/JHEP02(2018)141} {\bibfield  {journal}
  {\bibinfo  {journal} {JHEP}\ }\textbf {\bibinfo {volume} {02}},\ \bibinfo
  {pages} {141} (\bibinfo {year} {2018})},\ \Eprint
  {http://arxiv.org/abs/1801.00915} {arXiv:1801.00915 [hep-th]} \BibitemShut
  {NoStop}%
\bibitem [{\citenamefont {Ferrari}\ and\ \citenamefont
  {Quadri}(2004)}]{Ferrari:2004pd}%
  \BibitemOpen
  \bibfield  {author} {\bibinfo {author} {\bibfnamefont {R.}~\bibnamefont
  {Ferrari}}\ and\ \bibinfo {author} {\bibfnamefont {A.}~\bibnamefont
  {Quadri}},\ }\href {\doibase 10.1088/1126-6708/2004/11/019} {\bibfield
  {journal} {\bibinfo  {journal} {JHEP}\ }\textbf {\bibinfo {volume} {11}},\
  \bibinfo {pages} {019} (\bibinfo {year} {2004})},\ \Eprint
  {http://arxiv.org/abs/hep-th/0408168} {arXiv:hep-th/0408168 [hep-th]}
  \BibitemShut {NoStop}%
%%CITATION = HEP-TH/0408168;%%
\bibitem [{\citenamefont {Quadri}(2024)}]{Quadri:2024xuv}%
  \BibitemOpen
  \bibfield  {author} {\bibinfo {author} {\bibfnamefont {A.}~\bibnamefont
  {Quadri}},\ }\href@noop {} {\  (\bibinfo {year} {2024})},\ \Eprint
  {http://arxiv.org/abs/2401.00693} {arXiv:2401.00693 [hep-ph]} \BibitemShut
  {NoStop}%
\bibitem [{\citenamefont {Shtabovenko}\ \emph {et~al.}(2016)\citenamefont
  {Shtabovenko}, \citenamefont {Mertig},\ and\ \citenamefont
  {Orellana}}]{Shtabovenko:2016sxi}%
  \BibitemOpen
  \bibfield  {author} {\bibinfo {author} {\bibfnamefont {V.}~\bibnamefont
  {Shtabovenko}}, \bibinfo {author} {\bibfnamefont {R.}~\bibnamefont {Mertig}},
  \ and\ \bibinfo {author} {\bibfnamefont {F.}~\bibnamefont {Orellana}},\
  }\href {\doibase 10.1016/j.cpc.2016.06.008} {\bibfield  {journal} {\bibinfo
  {journal} {Comput. Phys. Commun.}\ }\textbf {\bibinfo {volume} {207}},\
  \bibinfo {pages} {432} (\bibinfo {year} {2016})},\ \Eprint
  {http://arxiv.org/abs/1601.01167} {arXiv:1601.01167 [hep-ph]} \BibitemShut
  {NoStop}%
\bibitem [{\citenamefont {Shtabovenko}\ \emph {et~al.}(2020)\citenamefont
  {Shtabovenko}, \citenamefont {Mertig},\ and\ \citenamefont
  {Orellana}}]{Shtabovenko:2020gxv}%
  \BibitemOpen
  \bibfield  {author} {\bibinfo {author} {\bibfnamefont {V.}~\bibnamefont
  {Shtabovenko}}, \bibinfo {author} {\bibfnamefont {R.}~\bibnamefont {Mertig}},
  \ and\ \bibinfo {author} {\bibfnamefont {F.}~\bibnamefont {Orellana}},\
  }\href {\doibase 10.1016/j.cpc.2020.107478} {\bibfield  {journal} {\bibinfo
  {journal} {Comput. Phys. Commun.}\ }\textbf {\bibinfo {volume} {256}},\
  \bibinfo {pages} {107478} (\bibinfo {year} {2020})},\ \Eprint
  {http://arxiv.org/abs/2001.04407} {arXiv:2001.04407 [hep-ph]} \BibitemShut
  {NoStop}%
\end{thebibliography}


\end{document}